\documentclass[useAMS,usenatbib]{mn2e}

\usepackage[pdftex]{graphicx} 
\usepackage{multirow} 
\usepackage[caption=false]{subfig} 
\usepackage{enumerate} 
\usepackage[T1]{fontenc} 

\def\reff@jnl#1{{\rm#1\/}}

\def\aj{\reff@jnl{AJ}}                  
\def\araa{\reff@jnl{ARA\&A}}            
\def\apj{\reff@jnl{ApJ}}                
\def\apjl{\reff@jnl{ApJ}}               
\def\apjs{\reff@jnl{ApJS}}              
\def\ao{\reff@jnl{Appl.Optics}}         
\def\apss{\reff@jnl{Ap\&SS}}            
\def\aap{\reff@jnl{A\&A}}               
\def\aapr{\reff@jnl{A\&A~Rev.}}         
\def\aaps{\reff@jnl{A\&AS}}             
\def\azh{\reff@jnl{AZh}}                        
\def\baas{\reff@jnl{BAAS}}              
\def\jrasc{\reff@jnl{JRASC}}            
\def\memras{\reff@jnl{MmRAS}}           
\def\mnras{\reff@jnl{MNRAS}}            
\def\pra{\reff@jnl{Phys.Rev.A}}         
\def\prb{\reff@jnl{Phys.Rev.B}}         
\def\prc{\reff@jnl{Phys.Rev.C}}         
\def\prd{\reff@jnl{Phys.Rev.D}}         
\def\prl{\reff@jnl{Phys.Rev.Lett}}      
\def\pasp{\reff@jnl{PASP}}              
\def\pasj{\reff@jnl{PASJ}}              
\def\qjras{\reff@jnl{QJRAS}}            
\def\skytel{\reff@jnl{S\&T}}            
\def\solphys{\reff@jnl{Solar~Phys.}}    
\def\sovast{\reff@jnl{Soviet~Ast.}}     
 \def\ssr{\reff@jnl{Space~Sci.Rev.}}     
\def\zap{\reff@jnl{ZAp}}                        
\def\nat{\reff@jnl{Nature}}             

\title[\textit{Spitzer} characterisation of dust in the Perseus cloud]{\textit{Spitzer} characterisation of dust in an anomalous emission region: the Perseus cloud}

\author[C. T. Tibbs et al.]{C.~T.~Tibbs,$\!^{1,2}$\thanks{E-mail: ctibbs@ipac.caltech.edu (CTT)} N.~Flagey,$\!^{2,3}$ R.~Paladini,$\!^{2}$ M.~Compi\`{e}gne,$\!^{2}$ S.~Shenoy,$\!^{4}$ S.~Carey,$\!^{2}$ \and  A.~Noriega-Crespo,$\!^{2}$ C.~Dickinson,$\!^{1}$ Y.~Ali-Ha{\"{\i}}moud,$\!^{5}$ S.~Casassus,$\!^{6}$ K.~Cleary,$\!^{5}$ \and R.~D.~Davies,$\!^{1}$ R.~J.~Davis,$\!^{1}$ C.~M.~Hirata,$\!^{5}$ R.~A.~Watson$^{1}$ \\
$^{1}$Jodrell Bank Centre for Astrophysics, School of Physics and Astronomy, The University of Manchester, Manchester, M13 9PL, UK \\
$^{2}$\textit{Spitzer} Science Center, California Institute of Technology, M/S 220-6, 1200 E. California Blvd., Pasadena, CA 91125, USA \\
$^{3}$Jet Propulsion Laboratory, California Institute of Technology, 4800 Oak Grove Drive, Pasadena, CA 91109, USA \\
$^{4}$Space Science Division, NASA Ames Research Center, M/S 245-6, Moffett Field, CA 94035, USA \\
$^{5}$Cahill Center for Astronomy and Astrophysics, California Institute of Technology, Pasadena, CA 91125, USA \\
$^{6}$Departamento de Astronom{\'{\i}}a, Universidad de Chile, Casilla 36-D, Santiago, Chile }

\begin{document}

\date{Accepted **insert**; Received **insert**}

\pagerange{\pageref{firstpage}--\pageref{lastpage}} 
\pubyear{}

\maketitle

\label{firstpage}


\begin{abstract}

Anomalous microwave emission is known to exist in the Perseus cloud. One of the most promising candidates to explain this excess of emission is electric dipole radiation from rapidly rotating very small dust grains, commonly referred to as spinning dust. Photometric data obtained with the \textit{Spitzer Space Telescope} have been reprocessed and used in conjunction with the dust emission model~\textsc{dustem} to characterise the properties of the dust within the cloud. This analysis has allowed us to constrain spatial variations in the strength of the interstellar radiation field~($\chi_\mathrm{ISRF}$), the mass abundances of the PAHs and VSGs relative to the BGs~(Y$_\mathrm{PAH}$ and Y$_\mathrm{VSG}$), the column density of hydrogen~(N$_\mathrm{H}$) and the equilibrium dust temperature~(T$_\mathrm{dust}$). The parameter maps of Y$_\mathrm{PAH}$, Y$_\mathrm{VSG}$ and $\chi_\mathrm{ISRF}$ are the first of their kind to be produced for the Perseus cloud, and we used these maps to investigate the physical conditions in which anomalous emission is observed. We find that in regions of anomalous emission the strength of the ISRF, and consequently the equilibrium temperature of the dust, is enhanced while there is no significant variation in the abundances of the PAHs and the VSGs or the column density of hydrogen. We interpret these results as an indication that the enhancement in $\chi_\mathrm{ISRF}$ might be affecting the properties of the small stochastically heated dust grains resulting in an increase in the spinning dust emission observed at 33~GHz. This is the first time that such an investigation has been performed, and we believe that this type of analysis creates a new perspective in the field of anomalous emission studies, and represents a powerful new tool for constraining spinning dust models.

\end{abstract}




\section{Introduction}

Cosmic Microwave Background~(CMB) anisotropy measurements have provided a great insight into the cosmological parameters that define our Universe. Obtaining these measurements to ever higher sensitivity is complicated by the presence of contaminating foregrounds, whose physical understanding is therefore critical. The discovery of a dust-correlated emission mechanism in the frequency range~10~--~100~GHz~\citep{Kogut:96, Leitch:97, deOC:97} has reignited the study of Galactic foregrounds as interstellar medium~(ISM) emission mechanisms. 

This dust-correlated emission, referred to as anomalous microwave emission or microwave excess, represents an additional component of the diffuse emission at microwave wavelengths. Most of our early knowledge of this type of emission has come from statistical analyses of large areas of sky \citep{deOC:02, Lagache:03, Davies:06, MD:08}. In recent years, a few detections from specific types of Galactic objects have been reported in the literature, such as H\textsc{ii} regions~\citep{Dickinson:07, Dickinson:09, Todorovic:10}, dark clouds~\citep{Finkbeiner:02, Casassus:06, Ami:09, Scaife:09, Dickinson:10, Vidal:11} and molecular clouds~\citep{Watson:05, Casassus:08, Tibbs:10}. 

Currently, the most popular hypothesis to explain the observed microwave excess is that of electric dipole radiation from rapidly rotating very small dust grains, also commonly referred to as `spinning dust'~\citep{DaL:98, Ali:09}. The spinning dust model generates a very distinctive peaked spectrum that rises between $\sim$~10~--~30~GHz, which cannot be easily reproduced by either free-free or synchrotron emission. Alternative models have been proposed, including hard synchrotron radiation \citep{Bennett:03} and magnetic dipole radiation \citep{DaL:99}, however, the spinning dust paradigm is the one that best reconciles model predictions with observations~\citep[e.g. see][]{Planck_Dickinson:11}.

Although the anomalous emission occurs in the microwave regime, the proposed spinning dust hypothesis implies that the origin of the emission is intrinsically associated with interstellar dust. A number of dust models have been postulated to explain the infrared~(IR) emission in the diffuse ISM~\citep{Desert:90, LiaD:01, DaLi:07, Compiegne:11}. Despite differences between models, it is generally accepted that the observed IR emission falls into two categories: 1) stochastic emission and 2) thermal equilibrium emission. The stochastic emission is attributed to polycyclic aromatic hydrocarbons~(PAHs) as well as to a population of small carbonaceous dust grains, while the thermal equilibrium emission is likely to be produced by large carbon and silicate dust grains. PAHs, first discovered by~\citet{Gillett:73}, are large molecules ($\sim$~1~nm in diameter) containing up to a few hundred carbon atoms. They are believed to be responsible for the broad emission features found in the near-IR spectrum~\citep{Leger:84}. The small carbonaceous dust grains are $\sim$~2~--~8~nm in diameter and thought to produce the mid-IR continuum emission. Finally, the large carbon and silicate grains, from a few nm to a few hundreds of nm in size, emit continuum radiation at far-IR wavelengths. Current spinning dust models~\citep{Ysard:10, Hoang:10, Silsbee:11} attribute the anomalous emission to the smallest dust grains.

\begin{figure}
\begin{center}
\includegraphics*[angle=0,scale=0.65,viewport=85 370 450 750]{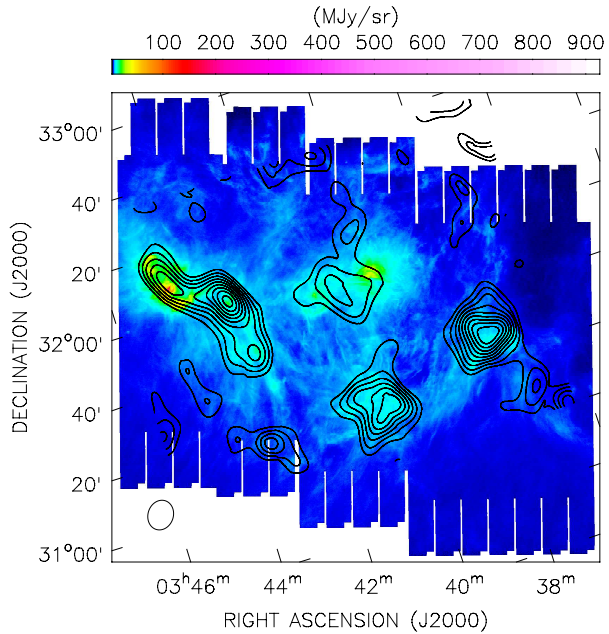}
\caption{Reprocessed \textit{Spitzer} MIPS 24~$\mu$m map of G159.6--18.5 overlaid with the VSA 33~GHz contours illustrating the high spatial correlation that exists between the microwave and IR emission. The bright central region of G159.6--18.5 corresponds to the B0~V star HD 278942, believed to be responsible for the H\textsc{ii} bubble. The VSA~$\approx$~7~arcmin FWHM synthesized beam is displayed in the bottom left-hand corner of the image, and the contours correspond to 10, 20, 30, 40, 50, 60, 70, 80 and 90~per cent of the peak flux, which is 200~mJy~beam$^{-1}$.}
\label{Fig:mips1-lgeom}
\end{center}
\end{figure}

Given the key role played by dust in the understanding of the microwave excess, an accurate characterisation of the dust properties in regions exhibiting anomalous emission is necessary to shed light on the nature of this effect. In this work we focus on the Perseus cloud, which to date is the most compelling example of anomalous emission associated with a specific Galactic region. The Perseus molecular cloud complex is a giant molecular cloud chain~$\sim$~30~pc in length, containing a total gas mass of~$\sim$~1.3$\times$10$^{4}$~M$_{\sun}$, and located at a distance of approximately 250~--~350~pc~\citep{Ridge:06a}. The Perseus cloud is interesting, and hence extensively studied, not only because of its relative proximity, but also because it forms low to intermediate mass stars, thus providing a link between the well known low mass star formation in the Taurus molecular cloud and the high mass star formation in the Orion molecular cloud. Within the cloud is the extended dust structure G159.6--18.5. Observations performed as part of the COMPLETE Survey~\citep{Ridge:06a} identified G159.6--18.5 as an expanding H\textsc{ii} bubble located on the far side of the cloud. \citet{Watson:05} observed G159.6--18.5 with the COSMOSOMAS experiment on degree angular scales, and found a rising spectrum between 11 and 16~GHz. Follow-up observations performed by~\citet{Tibbs:10} with the Very Small Array~(VSA) interferometer at 33~GHz on angular scales of~$\sim$~7~--~40~arcmin, identified five emission features displaying a remarkably high spatial correlation with the mid-IR emission~(see Fig.~\ref{Fig:mips1-lgeom}). The Perseus cloud has also been selected for extensive investigations of the microwave excess by the~\citet{Planck_Dickinson:11}. By combining the nine \textit{Planck} frequency bands in the range 28.5~--~857~GHz with ancillary data, and by subtracting the contaminating contributions arising from free-free emission, thermal dust emission and the CMB, the authors were able to produce an accurate and comprehensive spectrum of the anomalous emission on degree angular scales.

The layout of the paper is as follows. Section~\ref{sec:Spitzer} introduces the \textit{Spitzer} data used in this analysis and describes the data reprocessing. In Section~\ref{sec:Dustem} we combine the~\textit{Spitzer} data with a dust emission model~(\textsc{dustem}) to characterise the emission properties of the cloud. In Section~\ref{sec:spinning} we use the results of the dust modelling to characterise the spinning dust emission and we present our conclusions in Section~\ref{sec:conclusions}.


\section{\textit{Spitzer} Data}
\label{sec:Spitzer}


In order to characterise the dust properties across the Perseus cloud and investigate the link with the anomalous emission mechanism, we used multi-band IR data from the \textit{Spitzer Space Telescope}, taken as part of the `From Molecular Cores to Planet-Forming Disks'~\citep[c2d;][]{c2d:03} \textit{Spitzer} Legacy Program. The principal aim of the c2d program is to investigate the formation process of stars and planets, from prestellar cores to planet-forming disks. To achieve this goal, the c2d program involved both photometric and spectroscopic observations of five nearby molecular clouds, including the Perseus cloud. The photometric data are from the Multi-band Imaging Photometer for \textit{Spitzer}~(MIPS;~\citealt{Rieke:04}) and the Infrared Array Camera~(IRAC;~\citealt{Fazio:04}) while the spectroscopic observations were carried out with the Infrared Spectrograph~(IRS;~\citealt{Houck:04}). In the present paper, we limit our analysis to the MIPS and IRAC data, which provide a coverage of the Perseus cloud comparable to the VSA 33~GHz observations. 

Modelling the dust properties in the diffuse ISM depends critically on the quality of the IR observations. It is therefore crucial that we minimise any systematic effects that could potentially bias the analysis. Possible systematics in the \textit{Spitzer} data include stripes, gaps or artifacts, point sources and contamination from zodiacal light. Both the point sources and zodiacal light contamination are more pronounced at the shorter \textit{Spitzer} wavelengths. The density of sources increases with decreasing wavelength, while the zodiacal light peaks around 10~--~20~$\mu$m~\citep{Fixsen:02}, where its contribution can be comparable to the diffuse ISM emission.$\!$\footnote{For the Perseus cloud, it is of the order of 40 per cent at 8~$\mu$m.} For these reasons, we decided to reprocess both the IRAC and MIPS data. Such a reprocessing, in particular for the MIPS data, benefited from the techniques recently developed by the MIPSGAL (`A 24 and 70~$\mu$m Survey of the Inner Galactic Disk with MIPS') \textit{Spitzer} Legacy Program team~\citep{Carey:09}, which have been optimised for the treatment of extended emission in highly structured background regions such as the Galactic Plane and the Perseus cloud. 

We note that, for our reprocessing, we used, respectively, versions S16 and S14 of the \textit{Spitzer} Science Center~(SSC) pipeline for the MIPS and IRAC Basic Calibrated Data~(BCDs). The c2d analysis was instead carried out with version S13. The reprocessing of the photometric \textit{Spitzer} MIPS and IRAC data is described in Sections~\ref{sec:mips} and~\ref{sec:irac}, respectively.


\subsection{MIPS Data Reprocessing}
\label{sec:mips}

The MIPS instrument observes at three bands centred on 24, 70 and 160~$\mu$m, with angular resolutions of 6, 19 and 40~arcsec, respectively. Table~\ref{Table:MIPS_Summary} lists the MIPS data used in this analysis. Only the c2d Astronomical Observation Requests~(AORs) that provided coverage of the VSA 33~GHz observations were selected. The 24~$\mu$m data required minimal reprocessing, while the 70 and 160~$\mu$m data, had to be substantially reprocessed. The difference in data quality between 24~$\mu$m and 70 and 160~$\mu$m is caused by the different detector materials: the 24~$\mu$m detector is a Si:As array, while the 70 and 160~$\mu$m detectors are Ge:Ga arrays. In particular, the responsivity of the Ge:Ga arrays varies significantly with the build-up of charge from cosmic ray hits, while the Si:As array is less sensitive to this effect~\citep{Gordon:05}.


\subsubsection{24~$\mu$m Data}
\label{sec:24}

At 24~$\mu$m, all the MIPS first frames have a shorter exposure time. This results in a $\sim$~10~--~15~per cent suppression of the signal in the first frame. This effect also affects the second and third frames, although to a lesser extent, with a~$\sim$~2~per cent suppression. For this reason, the first three frames of each observation were discarded. This removal had no significant impact on the data as we had sufficiently many frames available to compensate. After discarding the first three frames, we subtracted the zodiacal light contribution from each BCD using the model by~\citet{Kelsall:98}, which is implemented in the SSC pipeline. This model is derived from observations with the Diffuse Infrared Background Experiment~(DIRBE) onboard the \textit{Cosmic Background Explorer}~(\textit{COBE}) satellite. Finally, an overlap correction routine, developed for the MIPSGAL survey~\citep{Mizuno:08}, was applied to match the background level of all the BCDs. The final mosaic was produced by combining the individual BCDs using the SSC MOsaicker and Point source EXtractor (MOPEX) software package.


\begin{table}
\begin{center}
\caption{Summary of the MIPS observations used in the present analysis. $^{\textit{a}}$~Units of Right Ascension are hours, minutes and seconds. $^{\textit{b}}$~Units of Declination are degrees, arcminutes and arcseconds.}
\begin{tabular}{c c c c c}
\hline
Field & RA$^{\textit{a}}$ & DEC$^{\textit{b}}$ & \multicolumn{2}{c}{AOR Key} \\
 & (J2000) & (J2000) & Epoch 1 & Epoch 2 \\
\hline
\hline
per2 & 03:44:30.7 & +32:06:08.1 & 5781248 & 5787904 \\
per3 & 03:42:34.5 & +31:55:33.0 & 5781504 & 5788160 \\
per4 & 03:40:39.0 & +31:37:53.0 & 5781760 & 5788416 \\
per5 & 03:37:42.0 & +31:16:41.0 & 5782016 & 5788672 \\
\hline
\label{Table:MIPS_Summary}
\end{tabular}
\end{center}
\end{table}


\subsubsection{70~$\mu$m Data}
\label{sec:70}

\begin{figure*}
\begin{center}$
\begin{array}{ccc}
\includegraphics*[angle=0,scale=0.45,viewport=85 370 440 780]{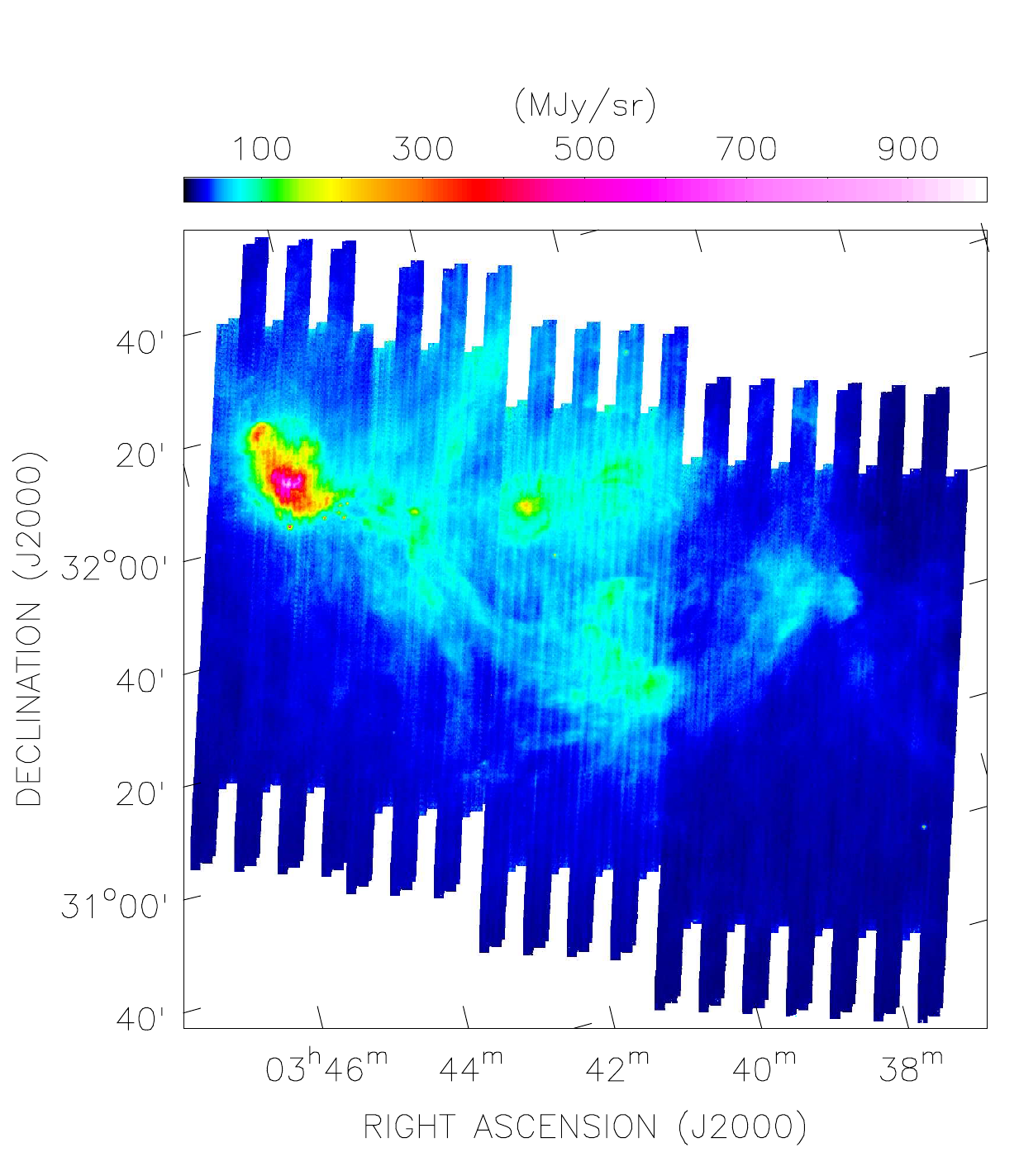} &
\includegraphics*[angle=0,scale=0.45,viewport=85 370 440 780]{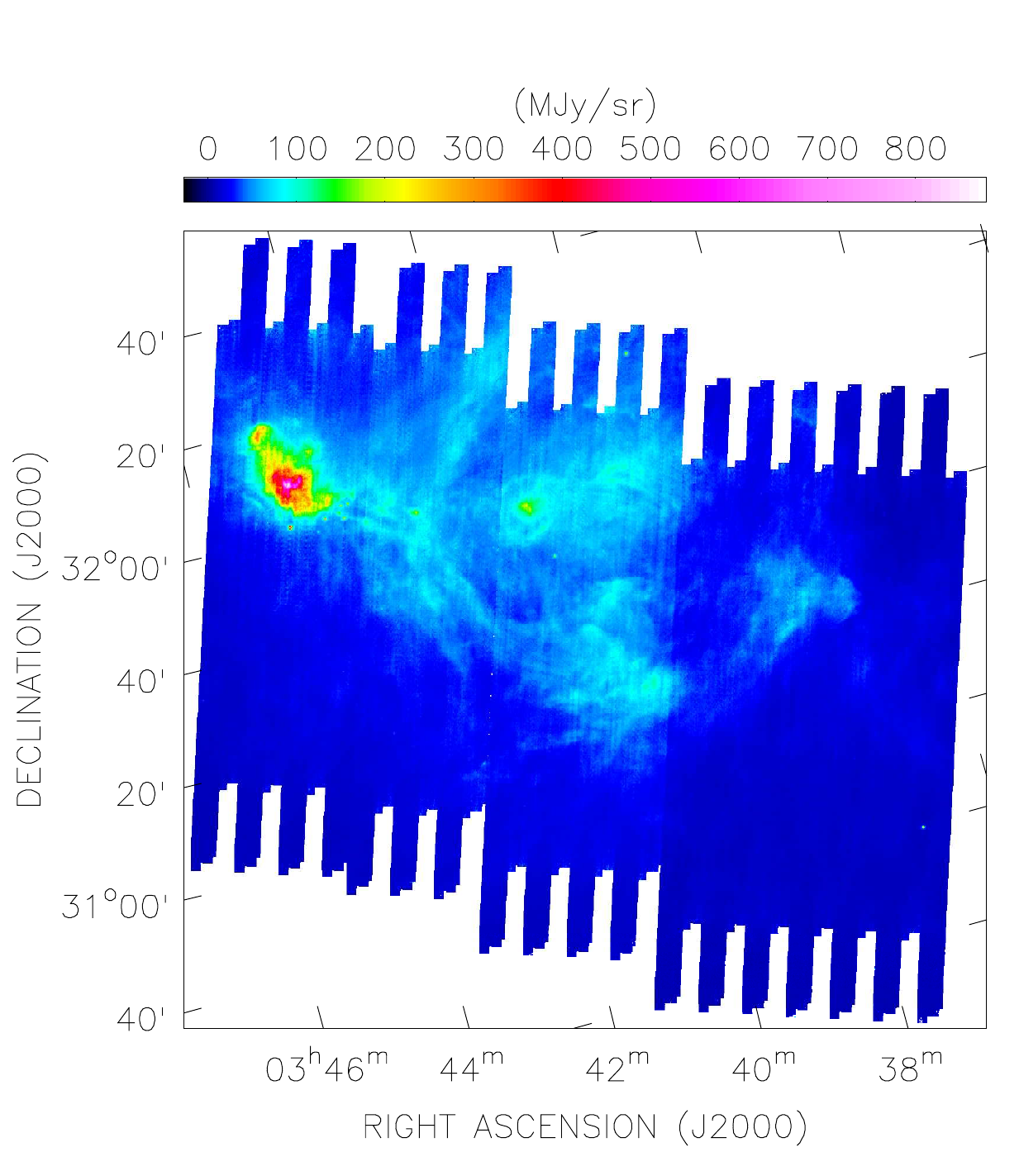} &
\includegraphics*[angle=0,scale=0.45,viewport=85 370 440 780]{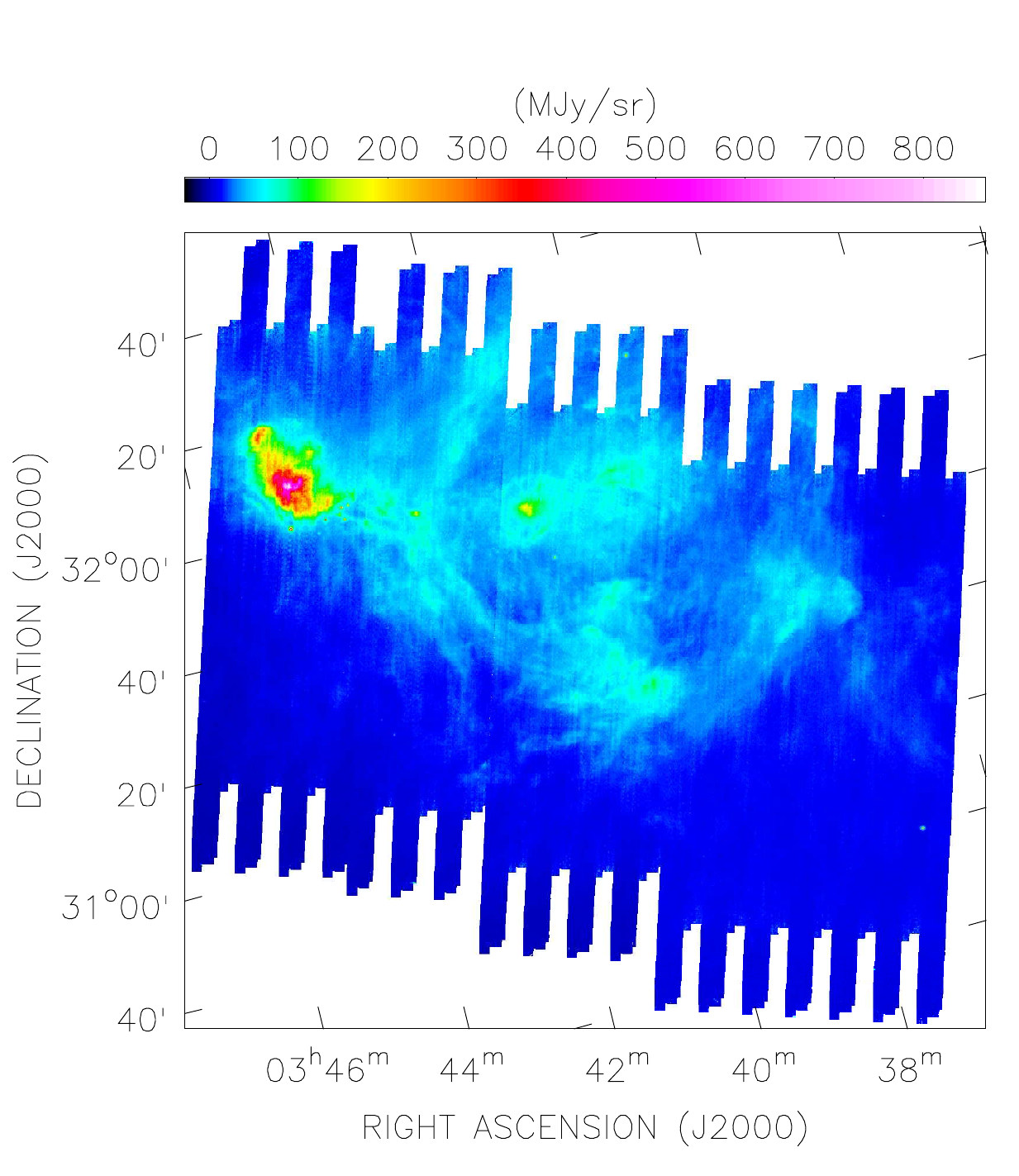} \\
\end{array}$
\caption{\textit{Spitzer} MIPS 70~$\mu$m maps: \textit{Left}: c2d map displaying the vertical striations caused by the variations in responsivity in the Ge:Ga detector and the artificial step caused by the thermal anneal; \textit{Middle}: Our map showing the result of running the GeRT software and the flat-fielding  which has removed the striations observed in the c2d map; \textit{Right}: Our map after correcting for the step caused by the thermal anneal. Source subtraction has not been performed on any of these images.}
\label{Fig:MIPS2}
\end{center}
\end{figure*}

For the MIPS 70~$\mu$m array, frequent ($\approx$~every 2 min) stimulator flashes provide a continuous measurement of the responsivity. However, if a stimulator flash occurs simultaneously on top of a bright region in the sky, it can yield an inaccurate estimate of the responsivity, and this in turn can have adverse effects, such as the production of streaks in the data. To remove artifacts due to responsivity drifts in our Ge:Ga data, we followed the procedure applied to the MIPSGAL 70~$\mu$m data as described in Paladini et al.~(in preparation).  The first step of this reprocessing makes use of the Germanium Reprocessing Tools (GeRT) software,$\!$\footnote{http://ssc.spitzer.caltech.edu/dataanalysistools/tools/gert/} one of the data reduction tools provided by the SSC. The GeRT software allows the user to run the Ge:Ga SSC pipeline offline, and customise its performance to his/her specific needs. Before running the GeRT software on our data, all the stimulator flash measurements were inspected and any erroneous measurements were removed using a moving median filter. The GeRT processing was then performed using a piecewise linear function to interpolate between the measurements. These newly created BCDs were then corrected with a flat-fielding routine to remove slowly varying gain fluctuations in the detector~(see~Fig.~\ref{Fig:MIPS2}).

\begin{figure*}
\begin{center}$
\begin{array}{ccc}
\includegraphics*[angle=0,scale=0.45,viewport=85 380 440 780]{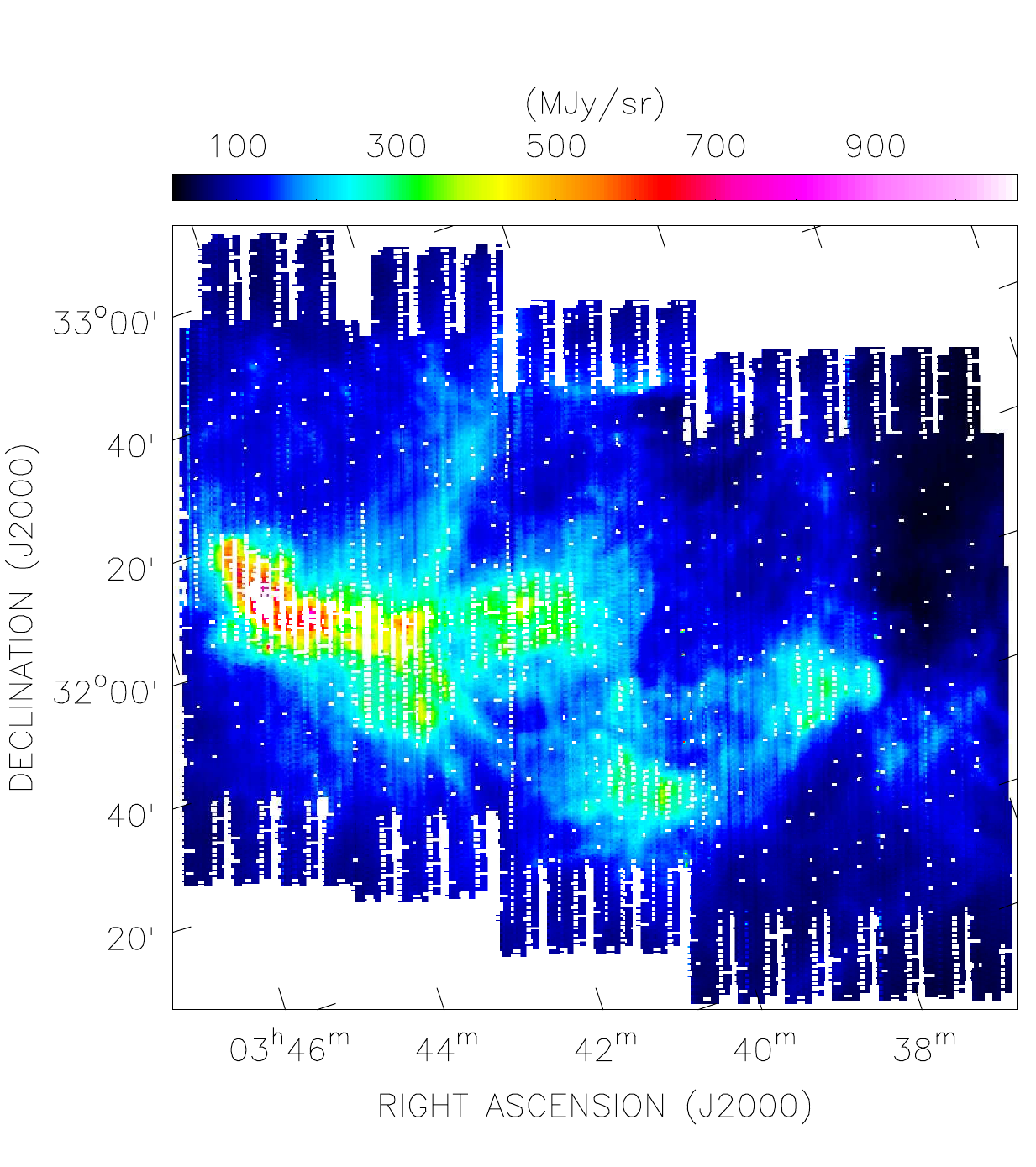} &
\includegraphics*[angle=0,scale=0.45,viewport=85 380 440 780]{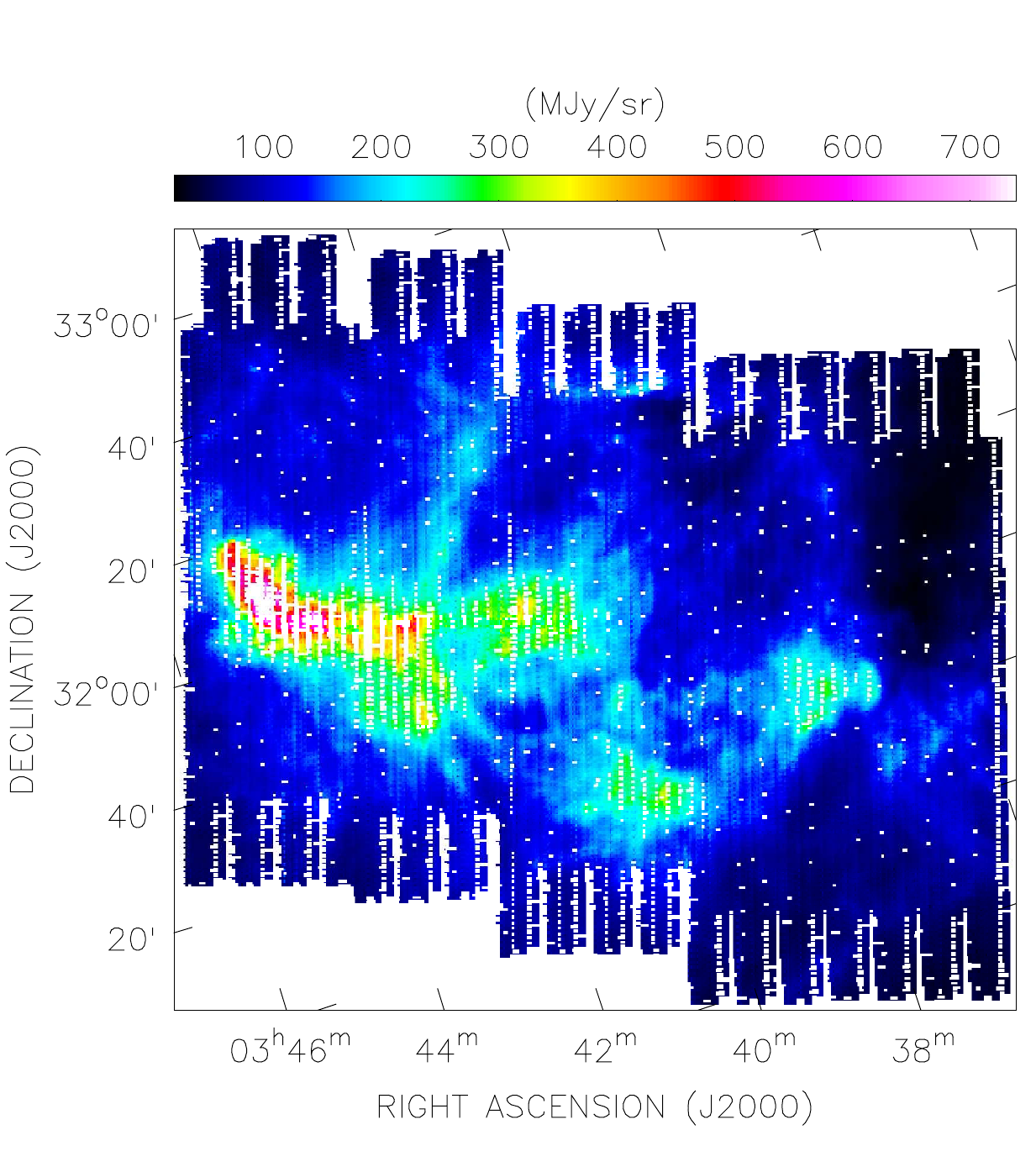} &
\includegraphics*[angle=0,scale=0.45,viewport=85 380 440 780]{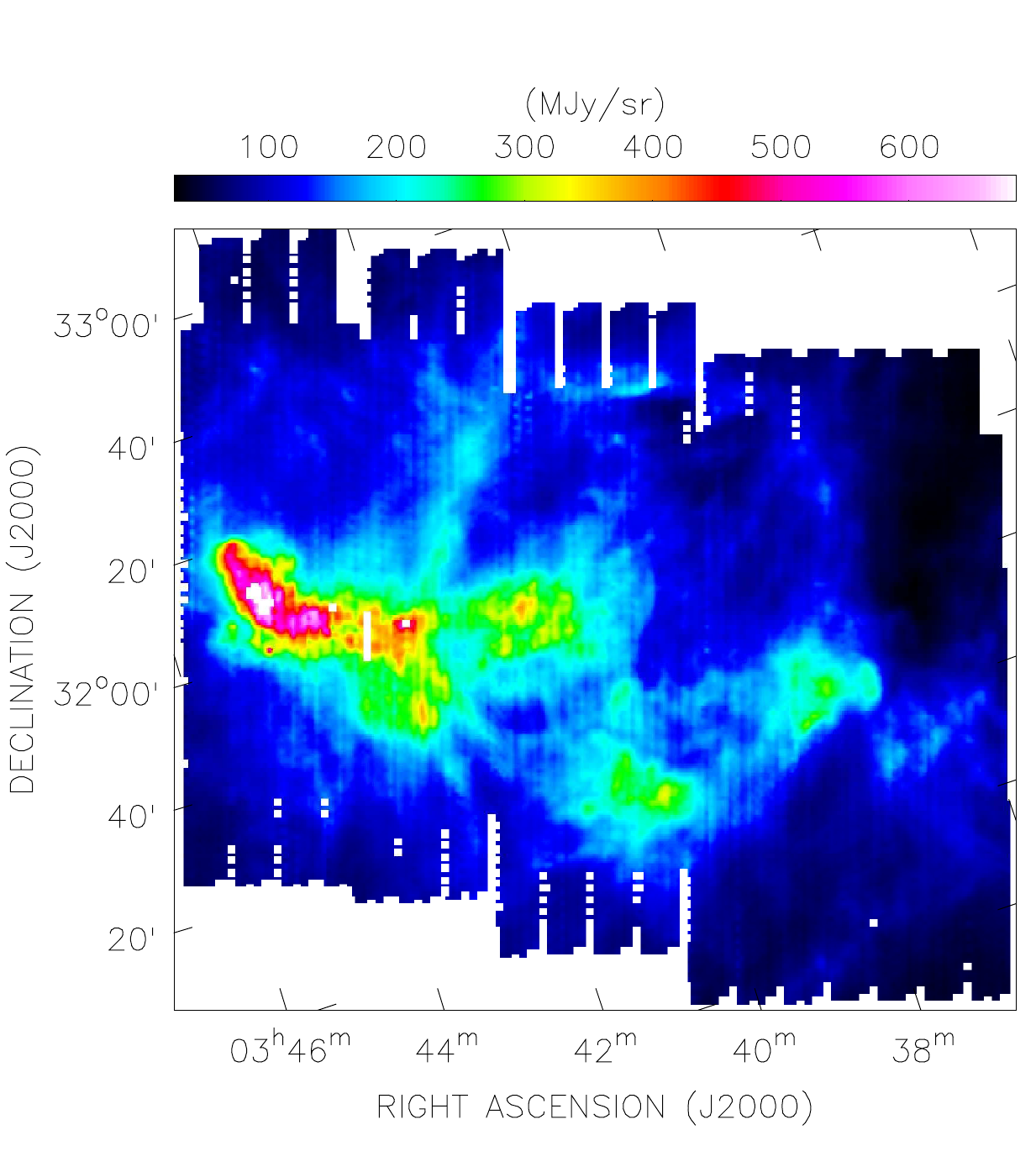} \\
\end{array}$
\caption{\textit{Spitzer} MIPS 160~$\mu$m maps: \textit{Left}: c2d map displaying the vertical striations caused by the variations in responsivity in the Ge:Ga detector and the gaps in the coverage; \textit{Middle}: Our map showing the result of running the GeRT software which has removed the striations observed in the c2d map; \textit{Right}: Our map after correcting for the lack of coverage using 40~arcsec pixels when creating the mosaic with MOPEX. Source subtraction has not been performed on any of these images.}
\label{Fig:MIPS3}
\end{center}
\end{figure*}

At the same time of the GeRT reprocessing and the flat-fielding, it was identified that a thermal anneal took place during observations of this region, creating an artificial step in the map~(see Fig.~\ref{Fig:MIPS2}). A thermal anneal occurs approximately every three hours, raising the temperature of the Ge:Ga detectors to $\approx$~8 and $\approx$~4~K for the 70 and 160~$\mu$m arrays, respectively, thereby erasing the accumulated effects of cosmic ray hits. To correct for the difference in responsivity before and after the anneal, we used IRIS~(Improved Reprocessing of the \textit{IRAS} Survey;~\citealt{IRIS:05}) data at 60~$\mu$m. This involved scaling the IRIS 60~$\mu$m map to 70~$\mu$m, assuming a standard SED for diffuse dust emission, and convolving the MIPS 70~$\mu$m map to the 4~arcmin angular resolution of the IRIS data. After performing a colour correction to account for the different bandpasses, we compared the IRIS and MIPS maps and found that the best match occurred after the thermal anneal had been performed. By performing five cuts across the step in the map, the median offset was found to be 10.3~MJy/sr, and the inaccurate BCDs were then corrected for this effect~(see Fig.~\ref{Fig:MIPS2}). As for the MIPS 24~$\mu$m data, the 70~$\mu$m data were then corrected for zodiacal light, although at this wavelength the effect is less significant. An overlap correction was performed, and finally, the corrected BCDs, ignoring stimulator flash frames, were combined with MOPEX to produce the final mosaic.


\subsubsection{160~$\mu$m Data}
\label{sec:160}

The 160~$\mu$m array, like the one at 70~$\mu$m, is a Ge:Ga detector. Hence, the 160~$\mu$m data had to be reprocessed using the GeRT software as described in Section~\ref{sec:70}. The newly produced BCDs did not, however, require flat-fielding, as the responsivity drift is in this case less severe than at 70~$\mu$m. At this wavelength the zodiacal light is negligible~($<$~2~per cent) and therefore its contribution did not have to be subtracted from the data. No overlap correction was required. 

The 160~$\mu$m detector has a much smaller field of view~(2.8~arcmin$^{2}$) compared with the 24 and 70~$\mu$m arrays~(29.2 and 13.7~arcmin$^{2}$, respectively). This, combined with the fact that over half of the detector has damaged pixels, results in an incomplete coverage of the map~(see Fig.~\ref{Fig:MIPS3}). To mitigate this effect, we slightly undersampled the image by selecting 40~arcsec pixels when producing the final mosaic, allowing MOPEX to interpolate over the gaps in the coverage. The map was then regridded from this coarser grid onto the native pixel grid~(see Fig.~\ref{Fig:MIPS3}).


\subsection{IRAC Data Reprocessing}
\label{sec:irac}

The IRAC instrument observes over the wavelength range 3.0~$\mu$m to 10.5~$\mu$m, in 4 bands centred on 3.6, 4.5, 5.8 and 8.0~$\mu$m, all with an angular resolution of~$\approx$~2~arcsec. Table~\ref{Table:IRAC_Summary} lists the IRAC data used in this analysis. Only the c2d AORs that provide coverage of the VSA 33~GHz observations were selected. IRAC observations are calibrated using point sources, but because we are interested in the diffuse emission, an extended emission correction was applied to the data, following the prescription described in the IRAC Data Handbook. All four IRAC bands suffer from well known artifacts such as muxbleed, column pulldown and electronic banding. To correct for these artifacts, as well as for the first frame effect, the IRAC BCDs were processed using contributed software provided by Sean Carey.$\!$\footnote{http://web.ipac.caltech.edu/staff/carey/irac\_artifacts/} After performing this artifact mitigation, the zodiacal light was subtracted. This subtraction was carried out individually on each BCD. It should be noted that when the IRAC observations were processed by the SSC pipeline, a dark field subtraction was performed, in order to correct for dark current and bias offsets. This sky dark observation also contains some contribution of zodiacal light, which then must be added back to the observations to avoid over-subtraction. To remove large scale gradients, a flat-fielding was performed. Finally an overlap correction was applied. As for the MIPS maps, the final mosaic was produced using MOPEX.


\begin{table}
\begin{center}
\caption{Summary of the IRAC observations used in the present analysis. $^{\textit{a}}$~Units of Right Ascension are hours, minutes and seconds. $^{\textit{b}}$~Units of Declination are degrees, arcminutes and arcseconds.}
 \begin{tabular}{c c c c c}
 \hline
  Field & RA$^{\textit{a}}$ & DEC$^{\textit{b}}$ & AOR Key \\
   & (J2000) & (J2000) & & \\
  \hline
  \hline
  per2 & 03:46:01.0 & +32:29:16.0 & 5790720 \\
  per4 & 03:44:53.3 & +31:39:00.0 & 5791232 \\
  per5 & 03:41:59.0 & +31:48:34.0 & 5791488 \\
  per6 & 03:40:12.0 & +31:26:41.0 & 5791744 \\
  per7 & 03:38:27.0 & +31:22:11.0 & 5792000 \\
  per8 & 03:36:23.0 & +31:08:41.0 & 5792256 \\
  per9 & 03:33:36.0 & +31:08:50.0 & 5792512 \\
  ic348 & 03:44:21.5 & +32:10:16.8 & 5790976 \\
  \hline
  \label{Table:IRAC_Summary}
\end{tabular}
\end{center}
\end{table}


\subsection{Point Source Extraction}
\label{subsec:source}

All \textit{Spitzer} bands, with the exception of MIPS 160~$\mu$m which is characterised by a very low source density, contain a significant number of point sources that require subtracting. The source extraction performed at all wavelengths~(excluding MIPS 160~$\mu$m for which no source subtraction was performed) made use of the Astronomical Point source Extractor~(APEX) tool, which is part of the SSC's MOPEX software. The APEX source detection and extraction consists of median filtering the background and subsequently detecting the sources using the Point Response Function (PRF). Following the recommendation by the SSC, for the MIPS channels we performed source subtraction on the final mosaics (single-frame mode), while for IRAC we first stacked all the BCDs and then detected the sources simultaneously in each frame (multi-frame mode). In addition, at 24 and 70~$\mu$m we built a customised PRF by selecting the brightest point-like sources in the cloud, while for the IRAC bands we used the in-flight IRAC PRFs, recently made available by the SSC.$\!$\footnote{http://ssc.spitzer.caltech.edu/irac/calibrationfiles/psfprf/}

For the MIPS 24~$\mu$m data, instead of using the standard APEX tool, we adopted the MIPSGAL point source extraction pipeline, which is optimised for detecting sources in regions of highly varying background, such as the Galactic Plane or the Perseus cloud. The MIPSGAL pipeline is a combination of APEX modules, \textsc{idl} routines and shell scripts, and consists of: 1) generating the core PRF by selecting sources in low background regions and with little or no structure; 2) combining this PRF with the wings of the theoretical PRF; 3) using the filtered PRF to extract the source flux and the unfiltered PRF to subtract the source from the input mosaic; 4) applying a scaling factor to correct the source flux for the median filtering. More details can be found in Shenoy et al.~(in preparation).


\subsection{Final Reprocessed Maps}
\label{subsec:final_maps}

The IRAC 3.6 and 4.5~$\mu$m mosaics are characterised by such a high density of sources that, even after performing source extraction, it makes them difficult to use to analyse extended emission. In addition to this, the contribution of stellar emission decreases, while the contribution of PAH emission increases, across the four IRAC bands, making IRAC band~4~(8~$\mu$m) the optimum band to trace the PAH emission. Also, the IRAC observations at all four channels were performed simultaneously, and due to the positioning of the IRAC detectors in the focal plane, the coverage common to both IRAC channels 3 and 4 is less than that of the individual channel coverage. For these reasons, we decided to ignore IRAC bands 1, 2 and 3 and use only IRAC band 4. We note that this channel, together with the three MIPS channels, provides coverage of both the stochastic and thermal equilibrium dust emission. In particular, the stochastic emission, produced by PAHs and small carbonaceous dust grains, will be dominant at 8 and 24~$\mu$m, while the thermal equilibrium dust emission, produced by the large carbon and silicate dust grains, will dominate at 160~$\mu$m. The emission at 70~$\mu$m is a result of both stochastic and thermal equilibrium dust emission~\citep{Compiegne:10}.

\begin{figure}
\begin{center}
\includegraphics*[angle=0,scale=0.55,viewport=75 450 480 760]{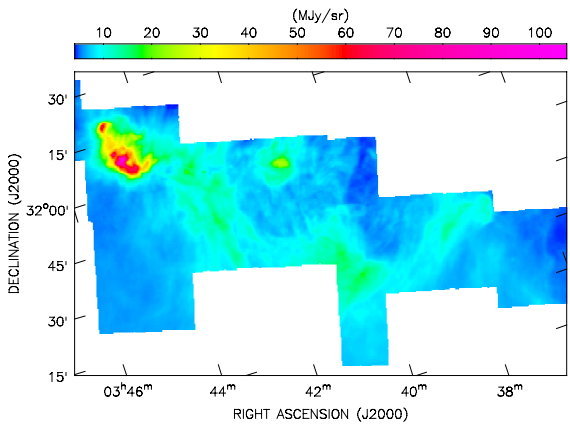} 
  \caption{Reprocessed \textit{Spitzer} IRAC 8~$\mu$m image at 40~arcsec angular resolution.}
     \label{Fig:Boxes}
   \end{center}
\end{figure}

The total coverage of our reprocessed maps is 2.3~deg$^{2}$ for the IRAC 8~$\mu$m map, 4.4~deg$^{2}$ for the MIPS 24~$\mu$m map, 3.9~deg$^{2}$ for the MIPS 70~$\mu$m map and 4.2~deg$^{2}$ for the MIPS 160~$\mu$m map. These maps are displayed in Figs.~\ref{Fig:Boxes},~\ref{Fig:mips1-lgeom},~\ref{Fig:MIPS2} and~\ref{Fig:MIPS3}, respectively. Conservative uncertainties for the flux in the \textit{Spitzer} maps, based on the intrinsic calibration of the instruments combined with the uncertainties introduced due to the data reprocessing, were estimated to be 5~per cent for the 8 and 24~$\mu$m maps, and 15~per cent for the 70 and 160~$\mu$m maps.


\section{Modelling the IR Dust Emission}
\label{sec:Dustem}

To investigate the interstellar dust we combined the \textit{Spitzer} data discussed in Section~\ref{sec:Spitzer} with the dust emission model \textsc{dustem}~\citep{Compiegne:11}. \textsc{dustem}, an updated version of the~\citet{Desert:90} model, allows the user to parameterise various dust emission and extinction properties. The software package incorporates five dust grain species: neutral and ionised PAHs~(log-normal size distribution with centre radius $a_{0}$~=~0.64~nm and width $\sigma$~=~0.1), small amorphous carbon grains~($a_{0}$~=~2.0~nm, $\sigma$~=~0.35), large amorphous carbon grains~(power-law size distribution with a minimum radius $a_{min}$~=~4.0~nm, a cut-off radius of $a_{t}$~=~150~nm, and a spectral index of $\alpha$~=~$-$2.8) and large silicate grains~($a_{min}$~=~4.0~nm, $a_{t}$~=~200~nm, $\alpha$~=~$-$3.4). For further details we refer the reader to~\citet{Compiegne:11}. For this analysis we combined the neutral and ionised PAHs and merged the large amorphous carbon and large silicate grains, creating three dust grain populations of PAHs, Very Small Grains~(VSGs) and Big Grains~(BGs).

\begin{figure*}
\begin{center}$
 \begin{array}{cc}
\includegraphics*[angle=0,scale=0.55,viewport=70 450 480 760]{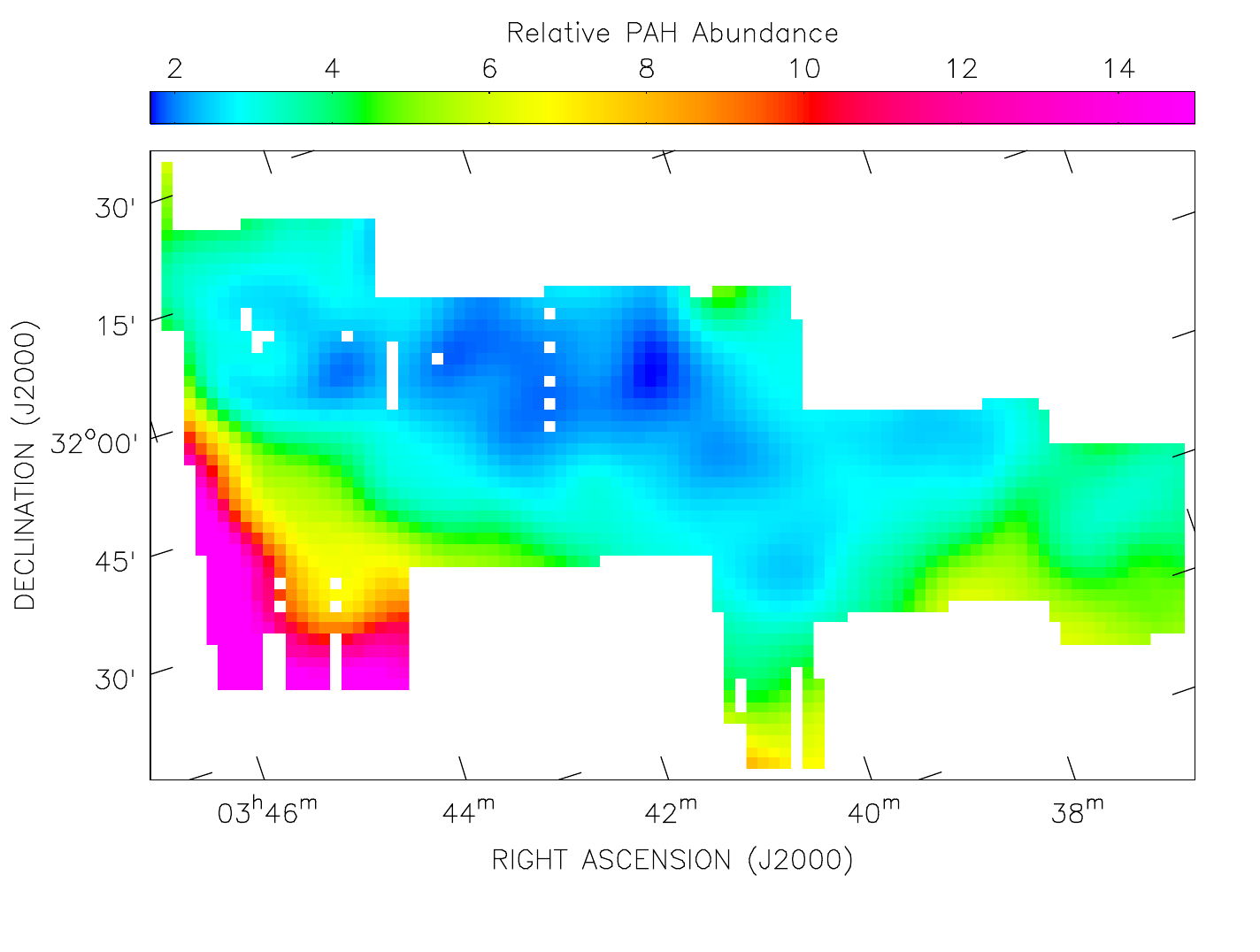} &
\includegraphics*[angle=0,scale=0.55,viewport=70 450 480 760]{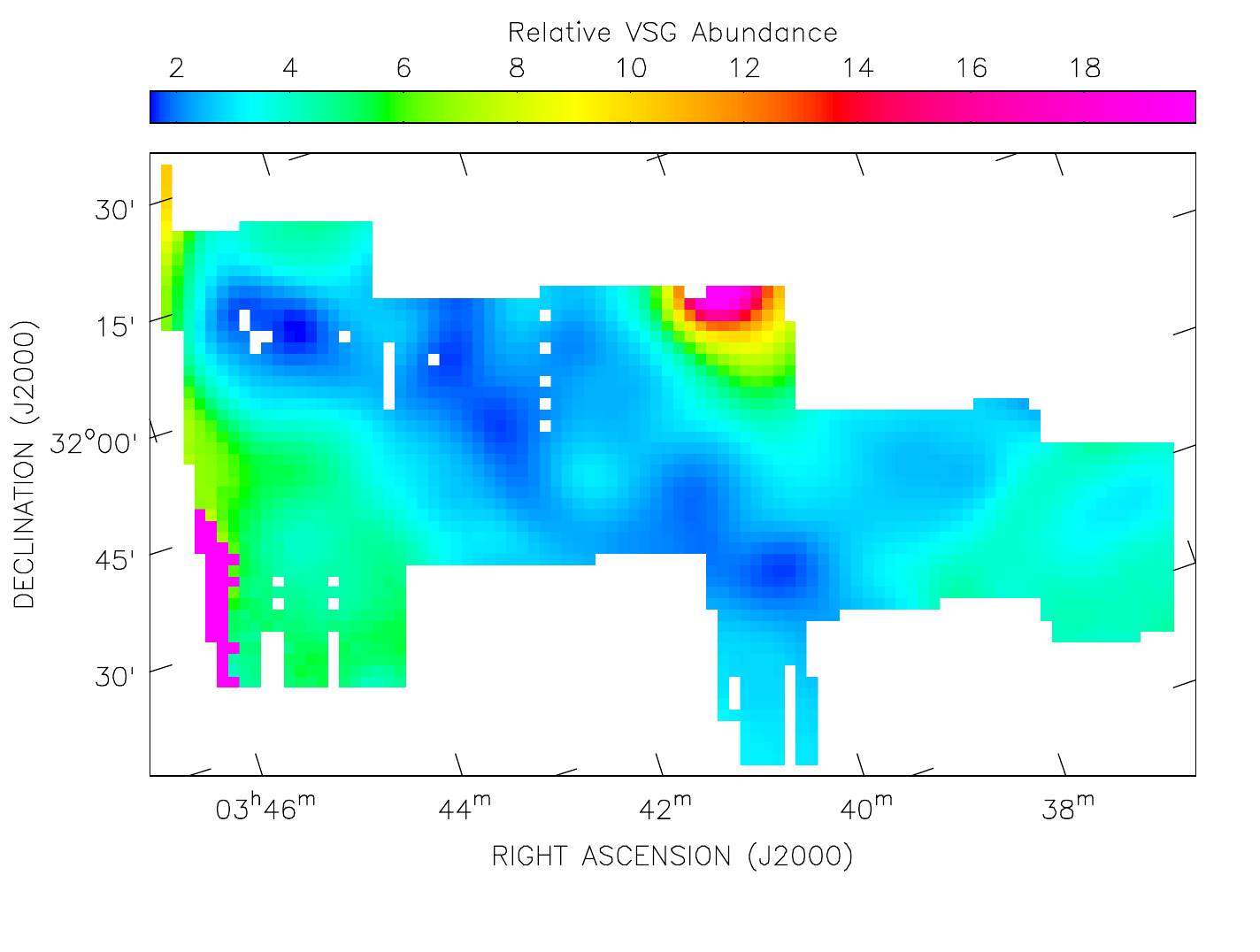} \\
\includegraphics*[angle=0,scale=0.55,viewport=70 450 480 760]{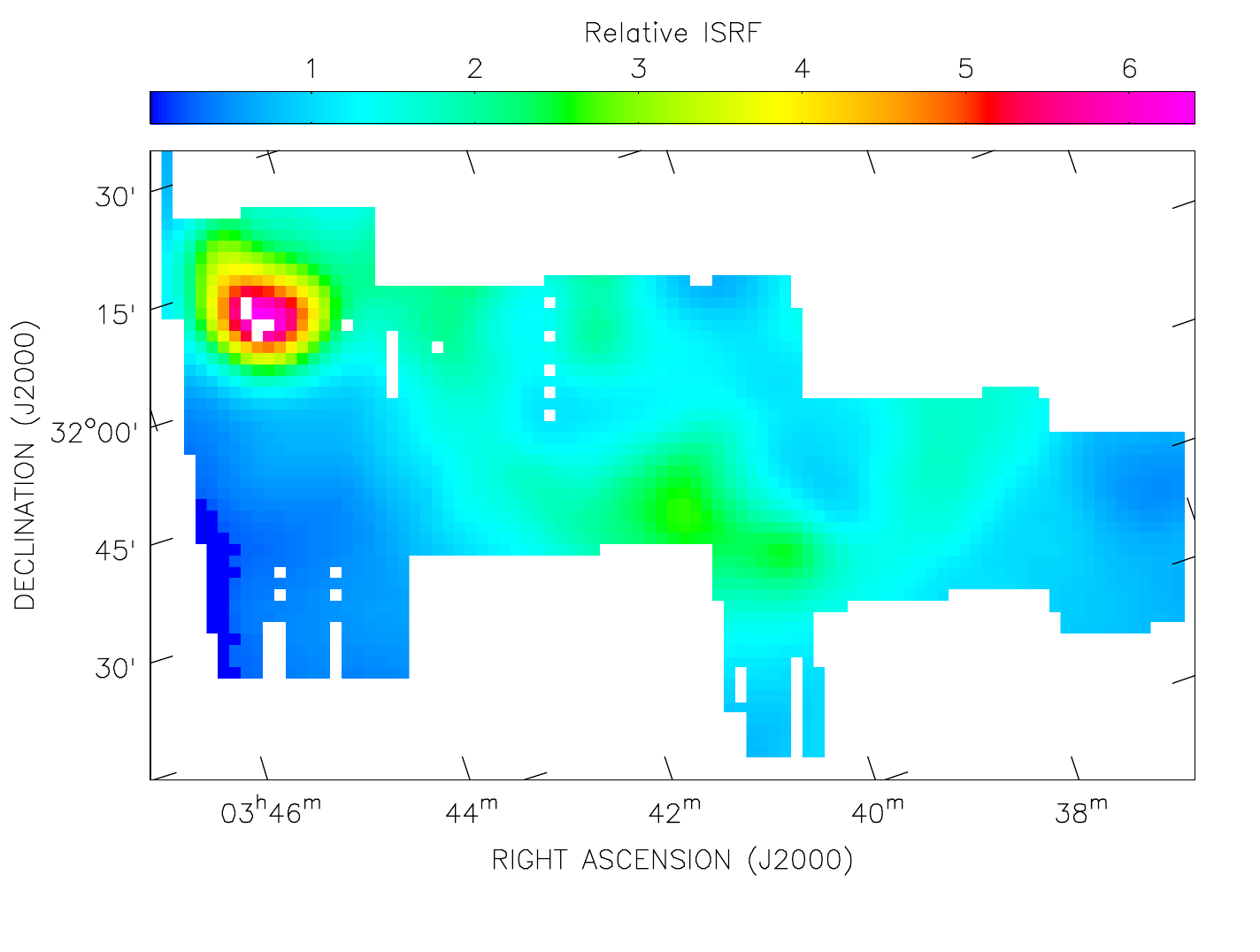} &
\includegraphics*[angle=0,scale=0.55,viewport=70 450 480 760]{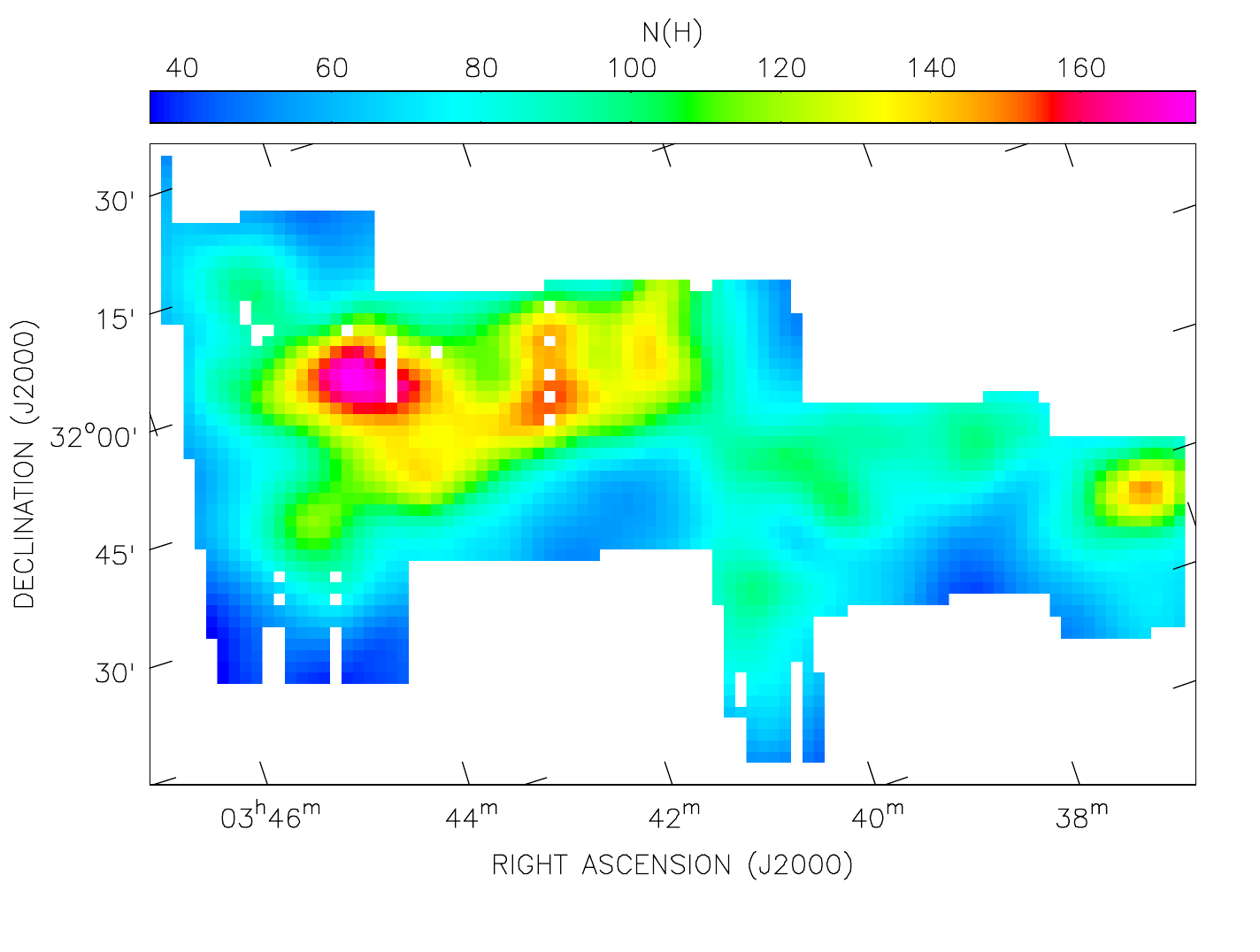} \\
\includegraphics*[angle=0,scale=0.55,viewport=70 450 480 760]{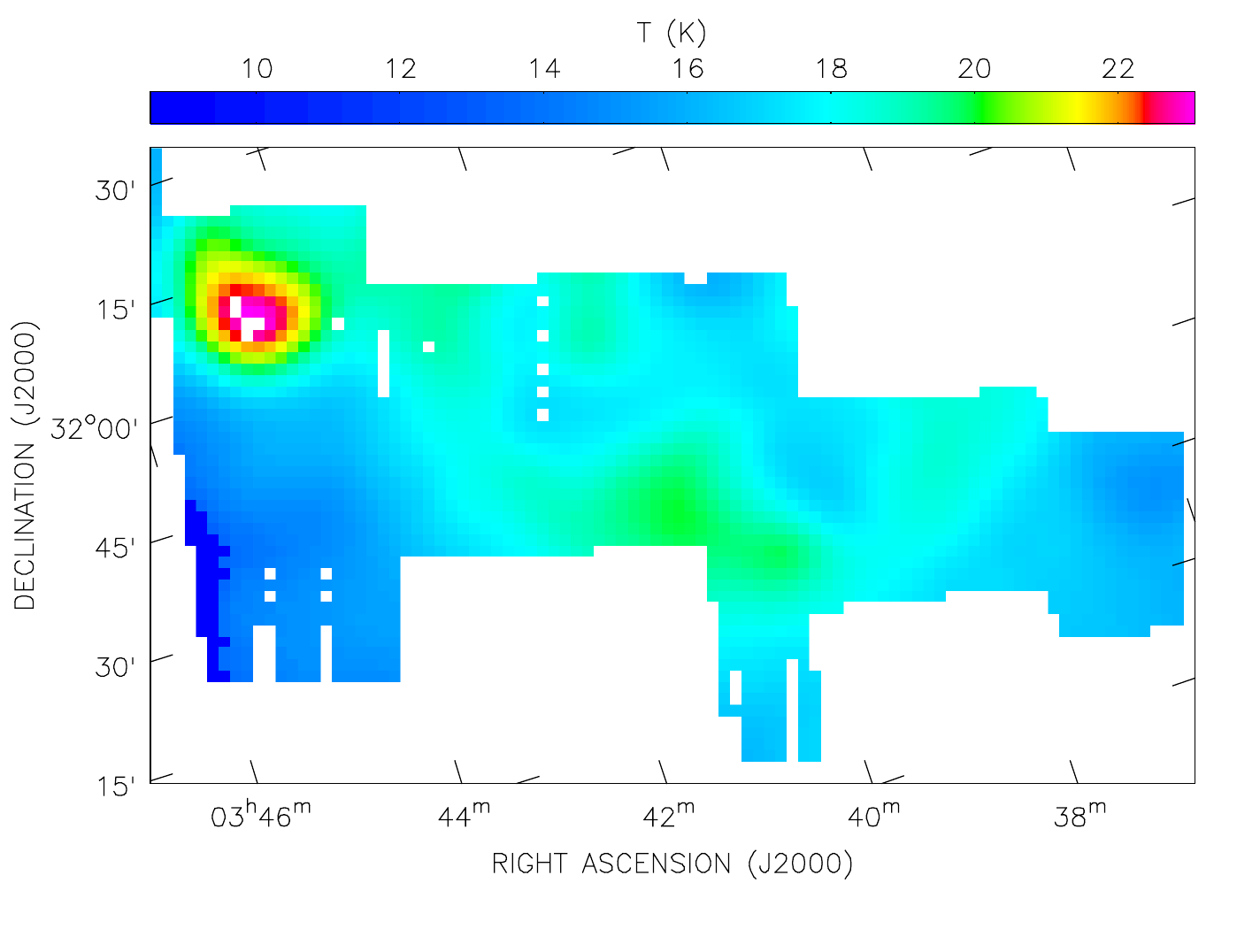} &
\includegraphics*[angle=0,scale=0.55,viewport=70 450 480 760]{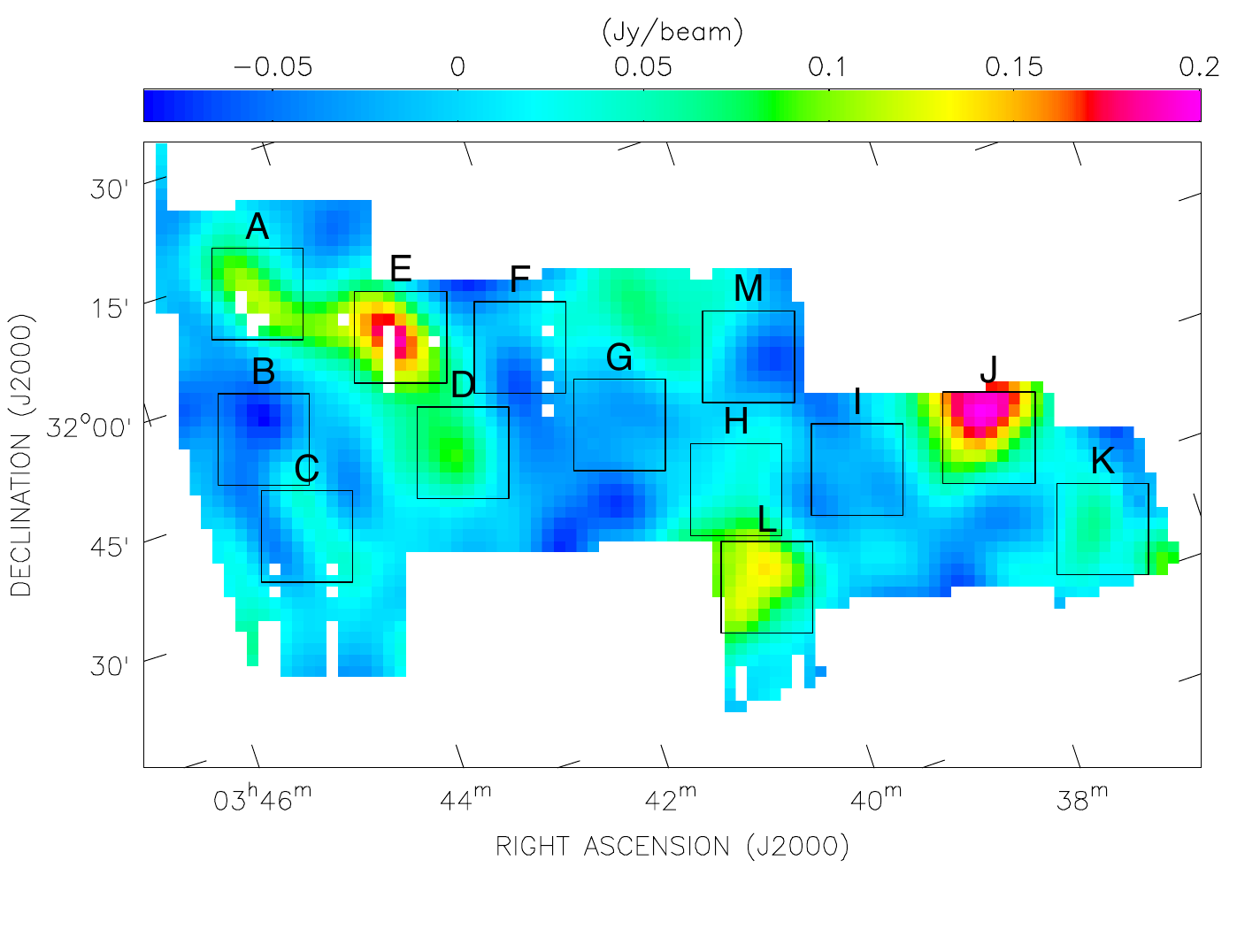} \\
  \end{array}$   
 \caption{Maps of the Perseus cloud: Y$_\mathrm{PAH}$ map~(\textit{Top Left}); Y$_\mathrm{VSG}$ map~(\textit{Top Right}); $\chi_\mathrm{ISRF}$ map~(\textit{Middle Left}); N$_\mathrm{H}$ map~(\textit{Middle Right}); T$_\mathrm{dust}$ map~(\textit{Bottom Left}); and the VSA 33~GHz map~(\textit{Bottom Right}).  The Y$_\mathrm{PAH}$, Y$_\mathrm{VSG}$ and $\chi_\mathrm{ISRF}$ parameter maps are the first to be produced for the Perseus cloud. Y$_\mathrm{PAH}$ and Y$_\mathrm{VSG}$ are given relative to the value for the diffuse high galactic latitude medium~\citep[see][]{Compiegne:11}, while $\chi_\mathrm{ISRF}$ is dimensionless. The N$_\mathrm{H}$ map was computed from the 2MASS/NICER extinction map and is in units of 10$^{20}$~H~cm$^{-2}$. The VSA map is provided to allow a visual comparison, and is overlaid with the thirteen regions listed in Table~\ref{Table:final_params}. All maps are displayed at 7~arcmin angular resolution.}
     \label{Fig:param_maps_7arc}
   \end{center}
\end{figure*}

\subsection{Constraining the IR Dust Emission}
\label{subsec:constraining}

To constrain the dust properties in the Perseus cloud we wrote an SED fitting routine, based on the \textsc{idl} routine \textsc{MPFIT}~\citep{Markwardt:09}, that combines the computed dust emissivities from \textsc{dustem} with the instrumental bandpass filters, and allows us to estimate the flux that would be observed by an instrument at a given wavelength. Using this fitting routine, we would like to constrain the mass abundances of the PAHs and VSGs relative to the BGs~(Y$_\mathrm{PAH}$ and Y$_\mathrm{VSG}$), the strength of the interstellar radiation field~(ISRF) relative to the~\citet{Mathis:83} solar neighbourhood value~($\chi_\mathrm{ISRF}$) and the column density of hydrogen~(N$_\mathrm{H}$). However, due to a lack of data at wavelengths longer than 160~$\mu$m, we do not measure the peak emission of the thermal equilibrium dust, which prevents us from recovering a reasonable estimate of the temperature and thus the strength of the ISRF. Consequently, there is a degeneracy between the strength of the ISRF and the column density of emitting material.
 
To circumvent this problem, we make use of the ancillary data from the COMPLETE survey~\citep{Ridge:06b}. This database is complementary to c2d, and includes extinction maps and emission maps of atomic and molecular line data. In particular, the COMPLETE extinction map of Perseus was generated by applying the NICER algorithm~\citep{Lombardi:01} to 2MASS data. Using N$_\mathrm{H}$/A$_\mathrm{V}$~=~1.87$\times$10$^{21}$~H~cm$^{-2}$~mag$^{-1}$ from~\citet{Bohlin:78}, we converted this extinction map into a map of the column density of hydrogen, hence breaking the degeneracy described above. Since the angular resolution of the 2MASS/NICER extinction map is 5~arcmin, this sets a lower limit to the angular resolution at which we can conduct our analysis. Taking into account that the VSA observations have an angular resolution of 7~arcmin, we degraded all four~\textit{Spitzer} maps, along with the 2MASS/NICER extinction map, to this resolution. After regridding them onto a 90~arcsec pixel grid, we incorporated the maps within our SED fitting routine. 

Having estimated N$_\mathrm{H}$, we keep this parameter fixed in the model, and we fit the 8, 24, 70 and 160~$\mu$m data points for Y$_\mathrm{PAH}$, Y$_\mathrm{VSG}$ and $\chi_\mathrm{ISRF}$, resulting in one degree of freedom~(\textit{dof}). We do not constrain the hardness of the ISRF, but make the assumption that it is similar to that of the solar neighbourhood. We note that if this is not the case, and if the ISRF is in fact harder in the Perseus cloud, then this would have the effect of decreasing the estimated abundances of both the PAHs and VSGs. Nevertheless, we believe that using the solar neighbourhood value is a reasonable approach in this case, given the lack of high-mass star formation in the cloud~\citep{Luhman:03}.

Running \textsc{dustem} and our SED fitting routine on a pixel by pixel basis resulted in a set of parameter maps~(Y$_\mathrm{PAH}$, Y$_\mathrm{VSG}$ and $\chi_\mathrm{ISRF}$) at 7~arcmin angular scales. In addition, for each pixel the SED fitting procedure also retrieves a temperature for each population of grains (PAHs, VSGs and BGs). These temperatures are not parameters in the fitting process, but are computed directly by \textsc{dustem} based on both the optical and thermal properties of the grains and the results of the fit. More precisely, given an exciting ISRF, \textsc{dustem} computes the temperature distribution for each size bin of each dust population taking into account the multi-photon absorption processes, based on the method described by~\citet{Desert:86}. Since only the BGs are in thermal equilibrium, we estimate the median for the BG size distribution only, and we take this value as representative of the dust equilibrium temperature in a given pixel (T$_\mathrm{dust}$). Fig.~\ref{Fig:param_maps_7arc} illustrates the parameter maps obtained for Y$_\mathrm{PAH}$, Y$_\mathrm{VSG}$, $\chi_\mathrm{ISRF}$ and T$_\mathrm{dust}$. For completeness, we also show the map of N$_\mathrm{H}$ derived from the COMPLETE extinction map. Y$_\mathrm{PAH}$ and Y$_\mathrm{VSG}$ are given relative to the value for the diffuse high galactic latitude medium~\citep[see][]{Compiegne:11}, $\chi_\mathrm{ISRF}$ is dimensionless and N$_\mathrm{H}$ is in units of 10$^{20}$~H~cm$^{-2}$. Remarkably, the maps of Y$_\mathrm{PAH}$, Y$_\mathrm{VSG}$ and $\chi_\mathrm{ISRF}$ are the first of their kind ever generated for the Perseus region, providing a comprehensive view of the dust temperature, radiation field and abundances of the different dust populations across the cloud. 

A careful inspection of these maps reveals that Y$_\mathrm{PAH}$ and Y$_\mathrm{VSG}$ both vary between~$\approx$~2 and 10 throughout the bulk of the cloud. N$_\mathrm{H}$ ranges between~$\approx$~4$\times$10$^{21}$ and~2$\times$10$^{22}$~H~cm$^{-2}$, and in regions where N$_\mathrm{H}$ is high~($\ge$~1$\times$10$^{22}$~H~cm$^{-2}$), both Y$_\mathrm{PAH}$ and Y$_\mathrm{VSG}$ appear to be decreased, suggesting that grain coagulation may be occurring~\citep[e.g.][]{Stepnik:03, Flagey:09}. $\chi_\mathrm{ISRF}$ ranges from~$\approx$~0.5~to~6, while the T$_\mathrm{dust}$ varies between~$\approx$~12~to~22~K throughout the cloud. In addition, $\chi_\mathrm{ISRF}$ and T$_\mathrm{dust}$ clearly trace the dust shell of G159.6--18.5, with both peaking around the open cluster IC~348. Finally, Y$_\mathrm{VSG}$ seems to be decreased in regions where $\chi_\mathrm{ISRF}$ is increased, which may indicate that the VSGs are being destroyed due to the increase in $\chi_\mathrm{ISRF}$.

\subsection{Investigating the Robustness of the Dust Modelling Analysis}
\label{subsec:robust}

To cross-check our results and investigate bias effects due to the adopted extinction map, we used two independent N$_\mathrm{H}$ constraints, in addition to the 2MASS/NICER map. Firstly we created an N$_\mathrm{H}$ map using the gas maps~(H\textsc{i} and $^{12}$CO) from the COMPLETE survey. The H\textsc{i} observations have an angular resolution of 7~arcmin and a spectral resolution of 0.32~km~s$^{-1}$. By integrating the H\textsc{i} data cube, assuming a spin temperature of 135~K~\citep{Dickey:78}, the H\textsc{i} data were converted into a H\textsc{i} column density map. The $^{12}$CO integrated map has an angular resolution of 46~arcsec and it was converted to a H$_{2}$ column density map using the standard relation X~=~N$_\mathrm{H_{2}}$/W($^{12}$CO), where W($^{12}$CO) is the integrated line intensity. We assumed a typical X factor value of 2$\times$10$^{20}$~cm$^{-2}$~K$^{-1}$~km$^{-1}$~s~\citep{Strong:96}. The total N$_\mathrm{H}$ map was produced by combining both maps~(N$_\mathrm{H_{\textsc{i}}}$ + 2$\times$N$_\mathrm{H_{2}}$) at a common angular resolution of 7~arcmin. Secondly, we used an extinction map of the region released as part of the c2d data products at an angular resolution of 5~arcmin. This map was generated in a similar manner to the 2MASS/NICER extinction map, however, in this case, both near- and mid-IR (1.25~--~24~$\mu$m) observations were folded in, allowing us to probe much higher extinction values~(up to A$_\mathrm{V}$~$\approx$~15~mag) than our original 2MASS/NICER map. This new extinction map was convolved to 7~arcmin. Both these new maps were used separately in our SED fitting routine to estimate N$_\mathrm{H}$ and both gave similar results to those obtained with the 2MASS/NICER extinction map. Due to the intrinsic variations in these three separate N$_\mathrm{H}$ constraints, the median variation in the parameters across the entire map was computed to be 5, 6, 13, 23 and 2~per cent for Y$_\mathrm{PAH}$, Y$_\mathrm{VSG}$, $\chi_\mathrm{ISRF}$, N$_\mathrm{H}$ and T$_\mathrm{dust}$, respectively. The large uncertainty on N$_\mathrm{H}$ is likely due to variations in the X factor across the cloud, as well as to the possible existence of a `dark gas' component~--~likely associated with a mixture of cold H\textsc{i} and warm H$_{2}$~--~that is not traced by either 21~cm measurements or by CO~line observations~\citep[see, for instance,][]{Planck_Bernard:11}. Having tested the N$_\mathrm{H}$ constraint and quantified the uncertainties, we combined them in quadrature with the fitting errors to estimate the final uncertainties on the parameters~(see Section~\ref{sec:spinning}).

\begin{figure}
\begin{center}
\includegraphics*[angle=180,scale=0.35,viewport=30 80 700 570]{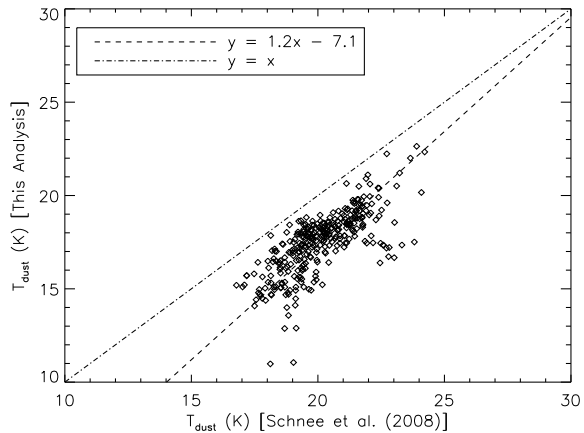} 
\caption{Scatter plot showing the comparison between the temperature map obtained in this analysis with \textsc{dustem} and the temperature map produced by~SGS08. The line of best fit is displayed~(\textit{dashed line}) along with the line $y=x$~(\textit{dash-dot line}).}
\label{Fig:temp_scatter}
\end{center}
\end{figure}

\begin{table*}
\begin{center}
\caption{Central coordinates of the thirteen 12~arcmin~$\times$~12~arcmin regions used in this analysis, and whether or not the region corresponds to an anomalous emission feature in the~\citet{Tibbs:10} VSA map. Also displayed are the mean parameters computed using \textsc{dustem} within each of the thirteen regions. $^{\textit{a}}$~Units of Right Ascension are hours, minutes and seconds. $^{\textit{b}}$~Units of Declination are degrees, arcminutes and arcseconds. $^{\textit{c}}$~Abundances relative to the BG abundance. $^{\textit{d}}$~Scaling factor relative to the \citet{Mathis:83} solar neighbourhood radiation field.}
\scalebox{1.0}{%
\begin{tabular}{c c c c c c c c c c}
\hline
Region & RA$^{\textit{a}}$ & DEC$^{\textit{b}}$ & Anomalous & Y$_\mathrm{PAH}$$^{\textit{c}}$ & Y$_\mathrm{VSG}$$^{\textit{c}}$ & $\chi_\mathrm{ISRF}$$^{\textit{d}}$ & T$_\mathrm{dust}$ & N$_\mathrm{H}$ & $\chi^{2}$/$\textit{dof}$ \\ 
 & (J2000) & (J2000) & Emission & & & & (K) & (10$^{20}$~H~cm$^{-2}$) & \\
\hline
\hline

A & 03:44:32 & +32:11:40 & Yes &       2.70 $\pm$       0.44 &       2.14 $\pm$       0.52 &       4.47 $\pm$       0.88 &       21.7 $\pm$        0.7 &       89.6 $\pm$       20.6 &        0.6 \\
B & 03:44:56 & +31:53:09 & No &       4.56 $\pm$       1.46 &       4.76 $\pm$       1.78 &       0.68 $\pm$       0.22 &       16.0 $\pm$        9.3 &       94.5 $\pm$       21.7 &       21.9 \\
C & 03:44:49 & +31:39:15 & No &       6.01 $\pm$       2.13 &       4.45 $\pm$       1.89 &       0.49 $\pm$       0.18 &       15.1 $\pm$        5.1 &      102.0 $\pm$       23.4 &       31.6 \\
D & 03:43:01 & +31:43:27 & Yes &       2.64 $\pm$       0.47 &       2.27 $\pm$       0.58 &       1.44 $\pm$       0.30 &       18.1 $\pm$        1.3 &      121.6 $\pm$       28.0 &        4.8 \\
E & 03:43:16 & +32:00:26 & Yes &       2.26 $\pm$       0.39 &       2.23 $\pm$       0.53 &       1.76 $\pm$       0.36 &       18.7 $\pm$        2.5 &      123.7 $\pm$       28.4 &        5.2 \\
F & 03:42:07 & +31:54:20 & No &       2.00 $\pm$       0.34 &       2.17 $\pm$       0.50 &       1.42 $\pm$       0.29 &       18.0 $\pm$        1.1 &      133.5 $\pm$       30.7 &        3.3 \\
G & 03:41:30 & +31:40:59 & No &       2.23 $\pm$       0.40 &       2.56 $\pm$       0.61 &       1.33 $\pm$       0.28 &       17.8 $\pm$        1.4 &      109.6 $\pm$       25.2 &        3.4 \\
H & 03:40:27 & +31:27:34 & Yes &       2.47 $\pm$       0.42 &       2.39 $\pm$       0.57 &       1.81 $\pm$       0.37 &       18.7 $\pm$        1.4 &       85.3 $\pm$       19.6 &        1.8 \\
I & 03:39:12 & +32:25:05 & No &       2.72 $\pm$       0.53 &       2.86 $\pm$       0.74 &       1.18 $\pm$       0.27 &       17.5 $\pm$        2.1 &       91.8 $\pm$       21.1 &        5.1 \\
J & 03:37:51 & +31:22:07 & Yes &       2.91 $\pm$       0.56 &       2.88 $\pm$       0.75 &       1.51 $\pm$       0.34 &       18.2 $\pm$        1.7 &       86.9 $\pm$       20.0 &        4.7 \\
K & 03:37:02 & +31:05:55 & No &       3.84 $\pm$       0.93 &       3.56 $\pm$       1.11 &       0.76 $\pm$       0.20 &       16.2 $\pm$        6.1 &       92.1 $\pm$       21.2 &       12.5 \\
L & 03:40:28 & +31:14:05 & Yes &       2.90 $\pm$       0.49 &       2.12 $\pm$       0.55 &       1.92 $\pm$       0.39 &       18.9 $\pm$        1.9 &       86.9 $\pm$       20.0 &        3.0 \\
M & 03:39:54 & +31:43:43 & No &       2.80 $\pm$       0.78 &       7.02 $\pm$       2.21 &       1.15 $\pm$       0.33 &       17.4 $\pm$        1.3 &       72.8 $\pm$       16.8 &        5.5 \\

\hline
\label{Table:final_params}
\end{tabular}}
\end{center}
\end{table*}

We also checked the reliability of the \textsc{dustem} analysis by comparing our temperature map with temperature maps previously obtained for this region. Temperature maps of the Perseus cloud have been created by \citet[][hereafter SGS08]{Schnee:08} who used various combinations of MIPS 70 and 160~$\mu$m and IRIS 60 and 100~$\mu$m maps to compute colour temperature maps. To compare their temperature map, and the one produced in this analysis, we convolved their 70/160 temperature map to 7~arcmin, regridded it onto our 90~arcsec pixel grid and trimmed it to match the coverage of our temperature map. A scatter plot of the two temperature maps is displayed in Fig.~\ref{Fig:temp_scatter} and we find that the temperature map computed with \textsc{dustem} is slightly lower~(by approximately 20 per cent) than the temperature map of~SGS08. There may be several reasons to explain this discrepancy. As discussed in Section~\ref{subsec:constraining}, the temperature computed by \textsc{dustem} is a physical temperature based on the optical and thermal properties of the grains, while the temperature produced by~SGS08 is a colour temperature based on a grey-body emission model. It is also possible that the median temperature we compute for the BG size distribution might not be completely representative of the effective temperature of the observed spectrum. Another possibility might be due to the fact that~SGS08 used the original c2d data in their analysis. In this work we have completely reprocessed the c2d data, as discussed in Section~\ref{sec:Spitzer}, and this is particularly important for the MIPS 70 and 160~$\mu$m maps due to the variations in the responsivity of the Ge:Ga detector. Last but not least, SGS08 might have under estimated the contribution of VSGs at 70~$\mu$m, which would result in a higher dust temperature estimate. Nevertheless, the two temperature maps show an overall correlation which, given that the two maps have been produced using significantly different methods, provides robustness to our analysis.


\section{Characterising the Anomalous Microwave Emission}
\label{sec:spinning}

The goal of this paper is to identify any potential correlations between spatial variations of the dust properties and the observed anomalous emission in the Perseus cloud. The parameter maps of Y$_\mathrm{PAH}$, Y$_\mathrm{VSG}$, $\chi_\mathrm{ISRF}$, N$_\mathrm{H}$ and T$_\mathrm{dust}$ provide a strong leverage to carry out such an investigation. For this purpose, we use the VSA 33 GHz map shown in Fig.~\ref{Fig:param_maps_7arc} as a reference. Since the VSA is an interferometer, the large scale structures are removed from the map due to incomplete sampling of the Fourier plane. However, as discussed in detail in~\citet{Tibbs:10}, one can clearly see that, on the spatial scales to which the instrument is indeed sentitive, some regions are characterised by the presence of anomalous emission, while others are not. We then use this information to investigate how the dust parameters in these regions compare with the parameters in regions where no anomalous emission has been detected. 

\begin{figure*}
\begin{center}$
 \begin{array}{cc}
\includegraphics*[angle=90,scale=0.35,viewport=60 10 550 700]{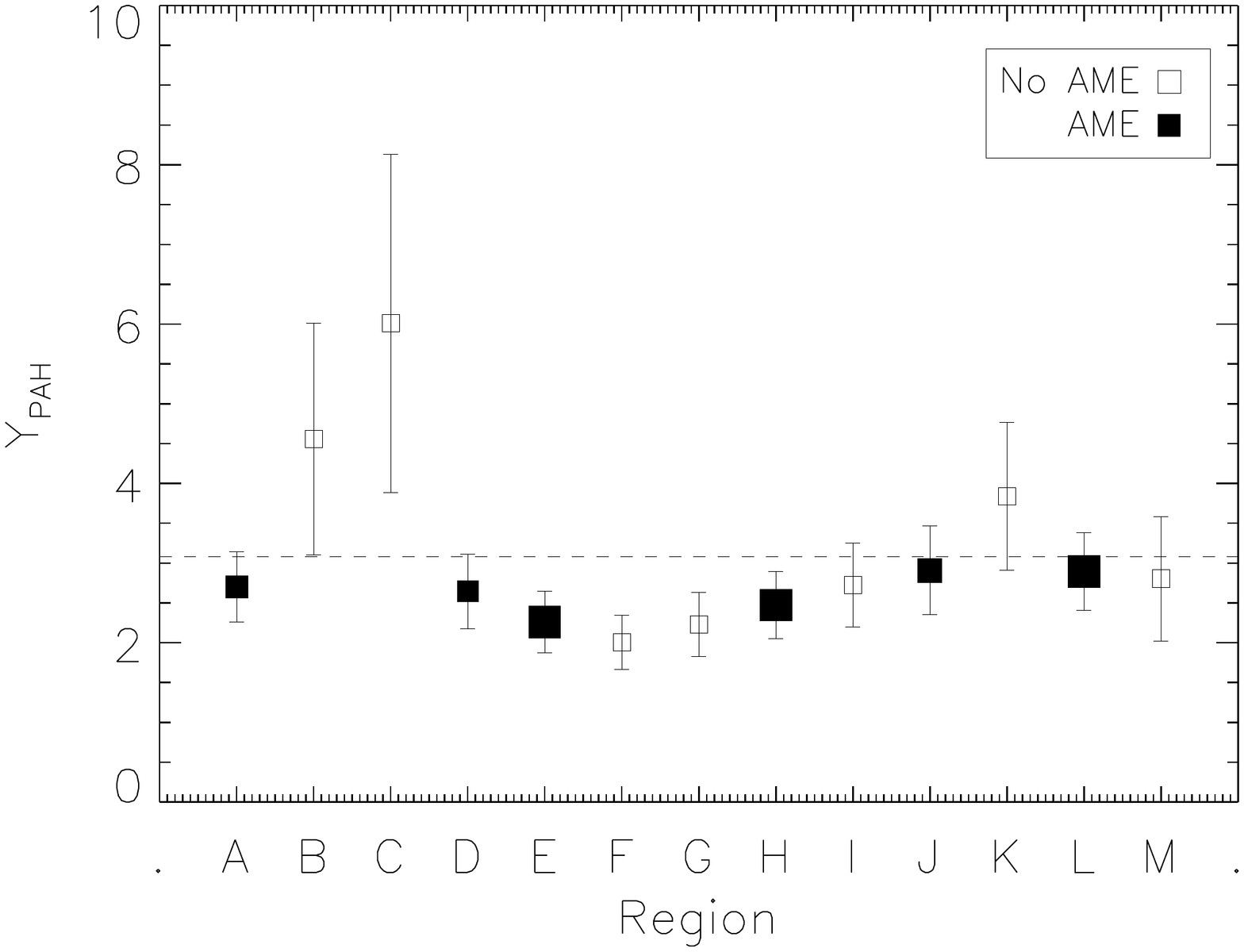} &
\includegraphics*[angle=90,scale=0.35,viewport=60 10 550 700]{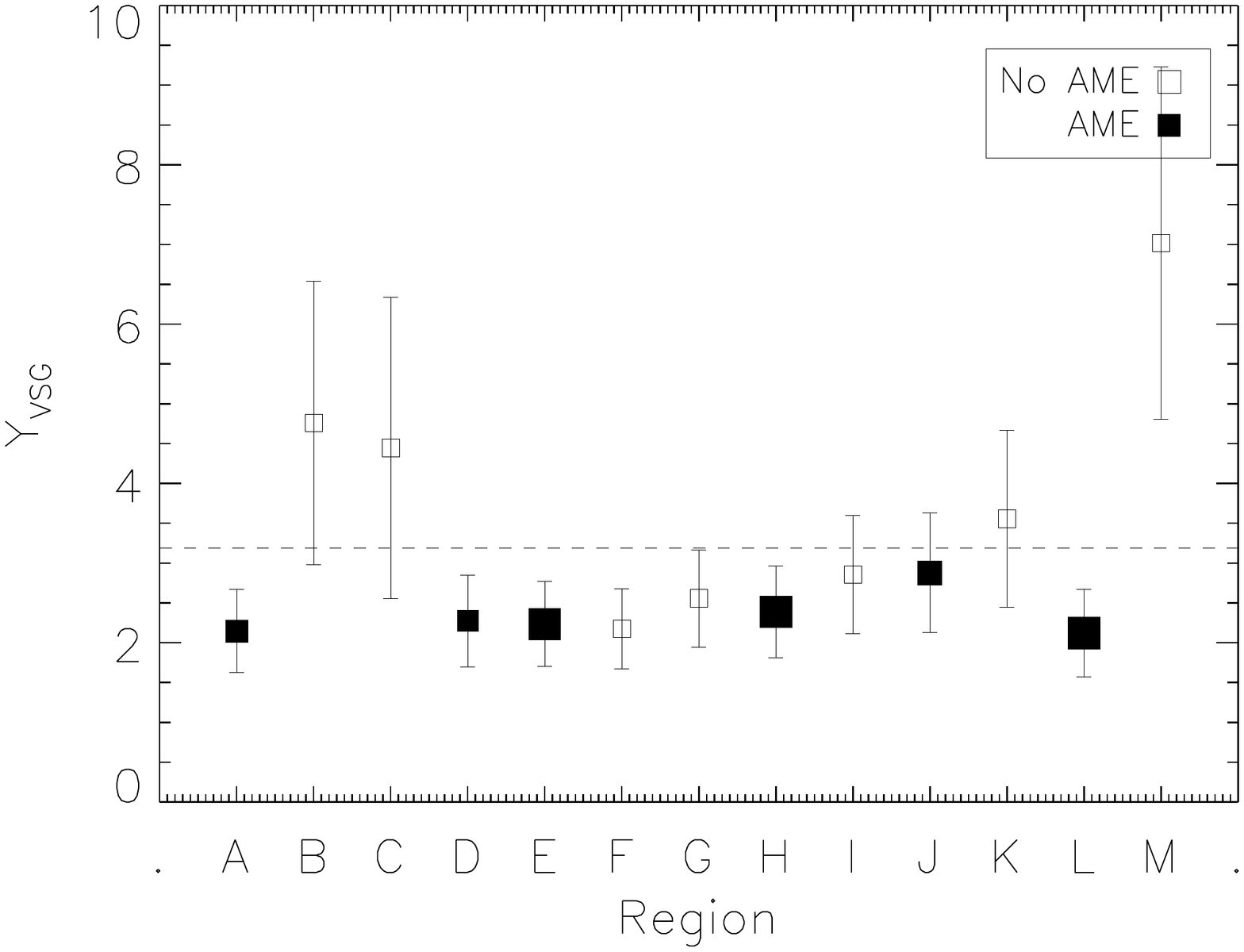} \\
\includegraphics*[angle=90,scale=0.35,viewport=60 10 550 700]{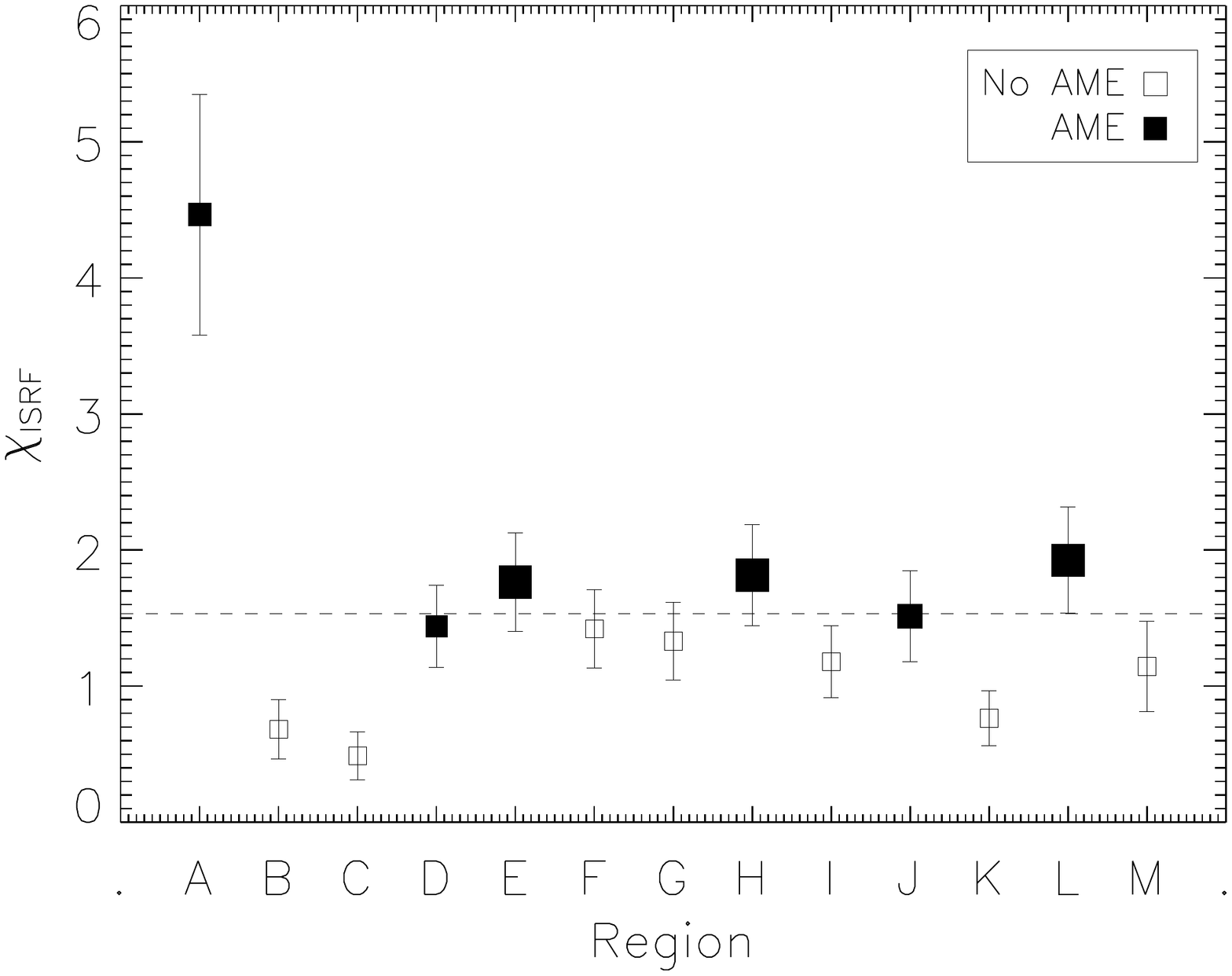} &
\includegraphics*[angle=90,scale=0.35,viewport=60 10 550 700]{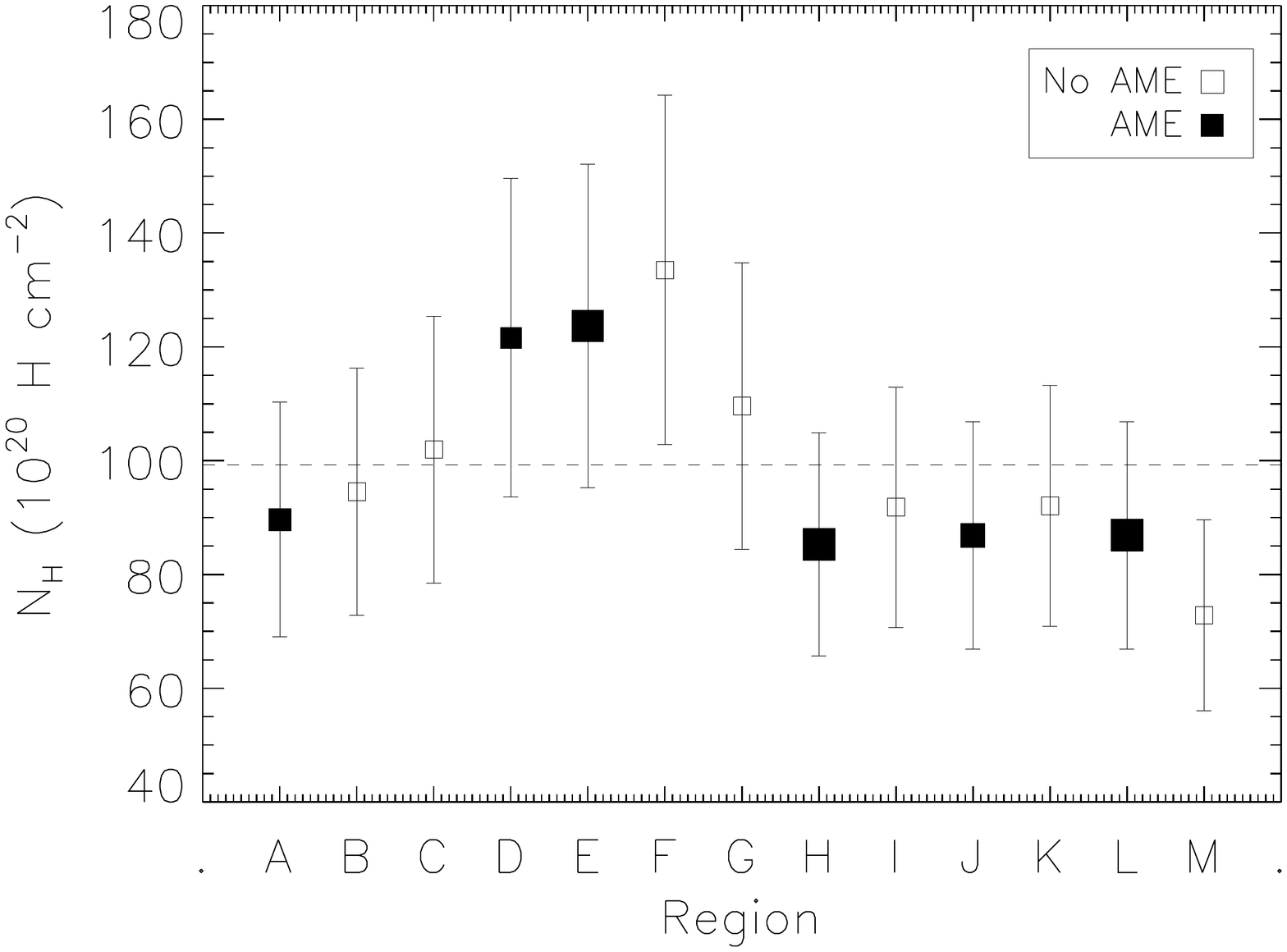} \\
\end{array}$
\includegraphics*[angle=90,scale=0.35,viewport=60 10 550 700]{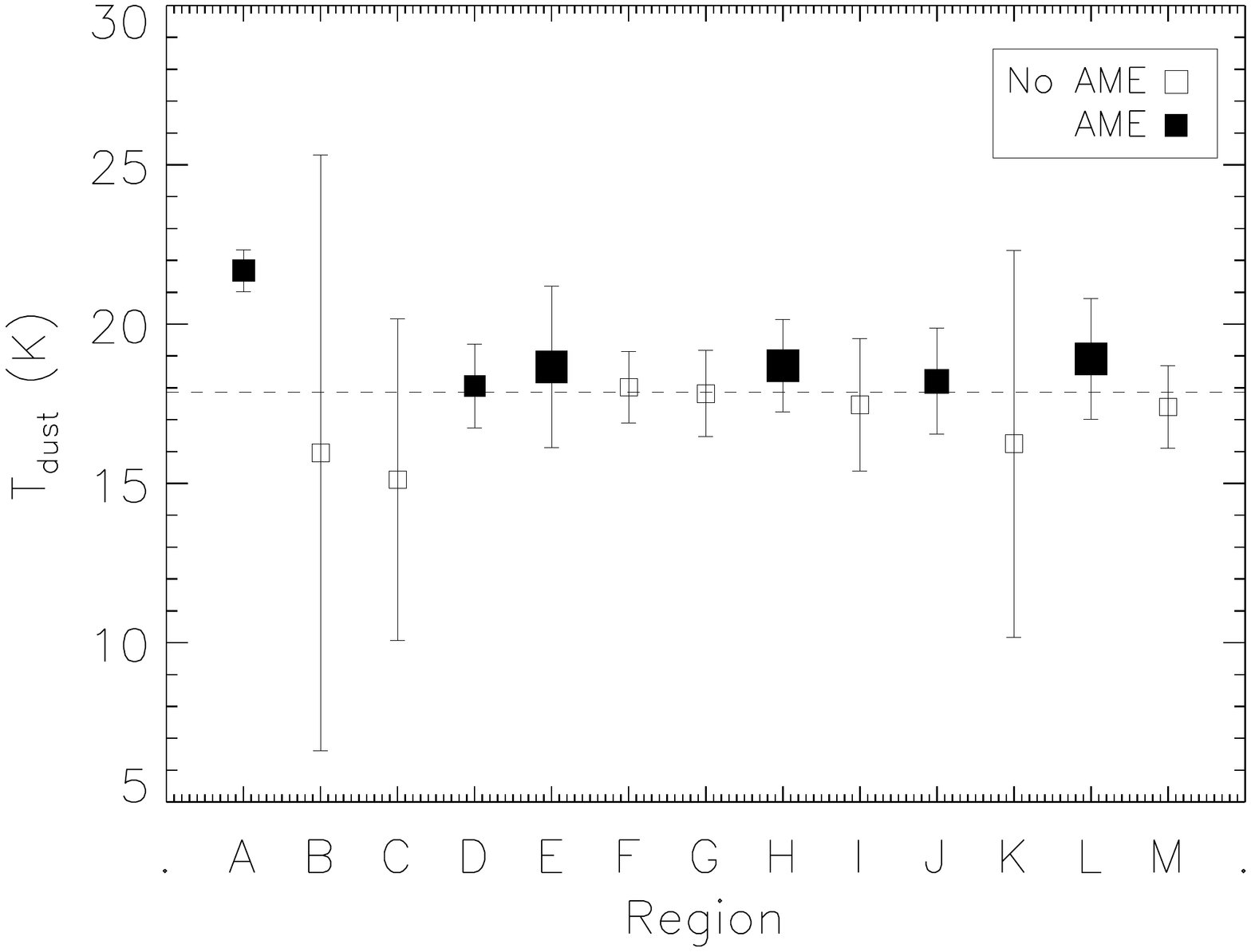} \\
  \caption{Plots displaying the computed parameters listed in Table~\ref{Table:final_params} for all thirteen regions, along with the mean value~(\textit{dashed line}). Regions with~(\textit{filled squares}) and without~(\textit{open squares}) anomalous emission are displayed, with the size of the squares representing the strength of the anomalous emission observed with the VSA~\citep{Tibbs:10}. These plots show that there appears to be a clear increase in both $\chi_\mathrm{ISRF}$ and T$_\mathrm{dust}$ in regions of anomalous emission, but that there does not appear to be any obvious correlation between the strength of the anomalous emission and the parameters computed in this analysis.}
     \label{Fig:final_params}
   \end{center}
\end{figure*}

For this purpose, thirteen regions were selected in which to perform a detailed analysis (see~Fig.~\ref{Fig:param_maps_7arc}). Each of these thirteen regions is 12~arcmin~$\times$~12~arcmin, providing coverage throughout the majority of the Perseus cloud. Table~\ref{Table:final_params} lists the central coordinates of the regions and whether or not the region corresponds to an anomalous emission feature in the VSA map. Also quoted in Table~\ref{Table:final_params}, and plotted in Fig.~\ref{Fig:final_params}, is the mean value of each of the parameters~(Y$_\mathrm{PAH}$, Y$_\mathrm{VSG}$, $\chi_\mathrm{ISRF}$, N$_\mathrm{H}$ and T$_\mathrm{dust}$) within each region. The mean value was computed assuming that all the pixels in each region are independent, and to assess whether the mean was a good statistical indicator in this case, we inspected the pixel value distribution for each region, and checked that it was approximately gaussian. The uncertainties on the parameters were estimated by combining the errors from the fit, which were computed from the covariance matrix, with the uncertainty due to using the 2MASS/NICER extinction map to constrain N$_\mathrm{H}$~(see Section~\ref{subsec:robust}). Additionally, Table~\ref{Table:final_params} lists the mean $\chi^{2}$/$\textit{dof}$ values obtained from the SED fitting procedure. The \textsc{dustem} fits appear to be reasonable for the majority of the regions, except for regions B, C and K~($\chi^{2}$/$\textit{dof}$~$>$~10). We believe the reason for this is that these regions are much colder, therefore the peak of the SED is at wavelengths considerably longer than 160~$\mu$m. As a consequence, the current set of data points allow us to only loosely constrain the IR emission in these regions. A better representation of their physical dust properties will be achieved by incorporating the~\textit{Herschel} SPIRE measurements, observed as part of the \textit{Herschel} Gould Belt Survey~\citep{Andre:10}, which will become publicly available at the end of the proprietary period.

By splitting the thirteen regions according to their location relative to the anomalous emission, a weighted mean was computed for each of the parameters listed in Table~\ref{Table:final_params}. The results of this analysis are illustrated in Table~\ref{Table:weighted_mean}. Both these results and Fig.~\ref{Fig:final_params} suggest a significant increase in $\chi_\mathrm{ISRF}$, and hence T$_\mathrm{dust}$, in regions of anomalous emission. In addition, we find that there is a slight decrease in the abundance of VSGs in regions of anomalous emission, however, this is only a 1.2$\sigma$ result and is therefore not statistically significant. We also find that there is no significant variation in the abundances of the PAHs or the column density of hydrogen between regions with and without anomalous emission. Interestingly, despite an enhancement in $\chi_\mathrm{ISRF}$ (and T$_\mathrm{dust}$) in regions of anomalous emission, there does not appear to be any obvious correlation between the strength of the anomalous emission observed with the VSA~(represented by the size of the squares in Fig.~\ref{Fig:final_params}) and these parameters. This is also true for the other parameters. This, and the fact that some of the results in Table~\ref{Table:weighted_mean} are only tentative, implies that anomalous emission is a very complex process to parameterise and is clearly dependent on a combination of multiple parameters, some of which we were not able to constrain in this analysis~(e.g. ionisation state and size distribution of the PAHs).


\begin{table}
\begin{center}
\caption{Results of performing a weighted mean analysis on the parameters in Table~\ref{Table:final_params}. These results show that for regions with anomalous emission there is a significant enhancement in the strength of the ISRF and the dust temperature.}
 \begin{tabular}{lccc}
 \hline
  Parameter & Anomalous & No Anomalous & Significance \\
   & Emission & Emission & \\
  \hline
  \hline
  
  Y$_\mathrm{PAH}$ & 2.60~$\pm$~0.18 & 2.44~$\pm$~0.21 & 0.6$\sigma$ \\ 
  Y$_\mathrm{VSG}$ & 2.29~$\pm$~0.23 & 2.75~$\pm$~0.31 & 1.2$\sigma$ \\      
  $\chi_\mathrm{ISRF}$ & 1.74~$\pm$~0.15 & 0.87~$\pm$~0.09 & 4.9$\sigma$ \\ 
  N$_\mathrm{H}$ & 94.3~$\pm$~9.0 & 93.8~$\pm$~8.3 & 0.0$\sigma$ \\  
  T$_\mathrm{dust}$ & 20.2~$\pm$~0.5 & 17.7~$\pm$~0.7 & 3.1$\sigma$ \\ 
  
  \hline
  \label{Table:weighted_mean}
\end{tabular}
\end{center}
\end{table}

The recent investigation by the~\citet{Planck_Dickinson:11} finds that the anomalous emission in the Perseus cloud can be accurately modelled by spinning dust emission arising from two gas components: a high density molecular gas component and a low density atomic gas component. The results of our analysis can also be explained in terms of the spinning dust paradigm. Since we find that there is no apparent significant variation in the abundances of PAHs and VSGs between regions with and without anomalous emission, the grain excitation mechanism will be the most important factor in producing spinning dust emission. The enhancement in $\chi_\mathrm{ISRF}$ in regions of anomalous emission will result in a higher density of photons which, in turn, will increase the photon~--~grain interactions, hence producing more excitation that could cause the dust grains to spin.

With the VSA,~\citet{Tibbs:10} found that the measurable microwave~--~IR correlation is restricted to the dense knots located along the shell of the H\textsc{ii} bubble G159.6--18.5 and the open cluster IC~348. It is known that the radiation pressure within expanding H\textsc{ii} regions sweeps up the surrounding dust forming a Photo Dissociation Region~(PDR;~\citealt[][]{Hollenbach:99}). This indicates that the knots of anomalous emission observed with the VSA along the shell are originating from the PDR surrounding the H\textsc{ii} gas, excited by the central B0~V star HD~278942. As an open cluster, we expect an increase in the strength of the ISRF observed towards IC~348 due to the presence of young stellar objects that are known to reside there~\citep[e.g.][]{Jorgensen:08}, thus possibly providing the source of excitation required to produce the observed anomalous emission in IC~348. Both of these scenarios are consistent with our results of finding an enhancement in $\chi_\mathrm{ISRF}$ in regions with anomalous emission, indicating that $\chi_\mathrm{ISRF}$, and hence photon~--~grain interactions, play a vital role in the excitation of the dust grains.

The \citet{Planck_Dickinson:11} results show in particular that the bulk of the anomalous emission is associated with the dense molecular gas component, and in such dense environments, gas~--~grain collisions are the predominant excitation mechanism of the dust grains. However, it must be remembered that the VSA is an interferometer, and as such it automatically filters out large scale structures, implying that the anomalous emission detected by the VSA is different to that detected by \textit{Planck}. In fact,~\citet{Tibbs:10} calculated that the VSA only observes approximately 10~per cent of the total 33~GHz emission while \textit{Planck} is sensitive to the entire cloud. Given that the star formation efficiency in the Perseus cloud is only 2.6~per cent~\citep{Jorgensen:08}, the bulk of the cloud would appear to be quiescent~--~a result which is consistent with the anomalous emission detected by \textit{Planck} being due to a dense molecular gas component. Moreover, the difference in the angular responses of the VSA and \textit{Planck} would seem to explain why $\chi_\mathrm{ISRF}$ appears to play an important role for the anomalous emission observed with the VSA, but not for the \textit{Planck} observations at degree angular scales.

Anomalous emission has been observed in a number of H\textsc{ii} regions~\citep[e.g.][]{Dickinson:07, Todorovic:10}, however, the observations lack the resolution to resolve their structure, thereby making it difficult to determine the exact location from which the emission is originating. G159.6--18.5 just happens to be large enough that the VSA observations can resolve the H\textsc{ii} region interior and identify that the anomalous emission is indeed produced in the PDR. The idea of anomalous emission originating from a PDR is not new, as~--~for instance~--~anomalous emission has previously been observed in a PDR in the $\rho$~Ophiuchi molecular cloud~\citep{Casassus:08}. 

\begin{figure}
\begin{center}
\includegraphics[angle=0,scale=0.37, viewport=78 50 900 550]{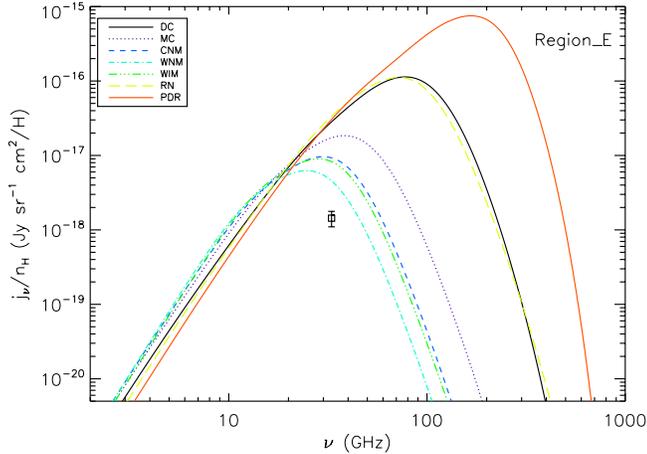} \\
\caption{Spinning dust curves computed using \textsc{spdust} with the $\chi_\mathrm{ISRF}$ values constrained with \textsc{dustem} for region E and idealised values for all other parameters. Also plotted is the spinning dust emissivity computed from the VSA observations. This shows that the modelled spinning dust curves are not consistent with the VSA observations, and that $\chi_\mathrm{ISRF}$ alone provides very little constraint on the computed spinning dust emission. This highlights the need for more constraint on the other \textsc{spdust} parameters.}
\label{Fig:spinning_dust}
\end{center}
\end{figure}

To further constrain the spinning dust emission model, we decided to combine the parameter values obtained with \textsc{dustem} with \textsc{spdust}~\citep{Ali:09, Silsbee:11}, a software package that computes a spinning dust spectrum for a set of user defined input parameters. Here we focus on $\chi_\mathrm{ISRF}$ as we have already shown that this parameter is enhanced in regions of anomalous emission. Taking the value of $\chi_\mathrm{ISRF}$ for each of the thirteen regions listed in Table~\ref{Table:final_params} and adopting all other \textsc{spdust} parameters used by the~\citet{Planck_Dickinson:11} for the Perseus cloud~(n$_\mathrm{H}$~=~250~cm$^{-3}$, T$_\mathrm{gas}$~=~40~K, n$_\mathrm{H^{+}}$/n$_\mathrm{H}$~=~1.12$\times$10$^{-4}$, n$_\mathrm{C^{+}}$/n$_\mathrm{H}$~=~1$\times$10$^{-4}$, 2n$_\mathrm{H_{2}}$/n$_\mathrm{H}$~=~1), we estimated the corresponding spinning dust curves. Furthermore, we used our values of $\chi_\mathrm{ISRF}$ in combination with the idealised environmental parameters for other phases of the ISM derived by~\citet{DaL:98}, to compute spinning dust curves for a Dark Cloud~(DC), Cold Neutral Medium~(CNM), Warm Neutral Medium~(WNM), Warm Ionised Medium~(WIM), Reflection Nebula~(RN) and~PDR. In this work we used the updated version of \textsc{spdust} as described by~\citet{Silsbee:11}, which assumes that the grains are randomly orientated with respect to their angular momentum axis. The computed spinning dust curves for one of the regions with anomalous emission~(region E) are displayed in Fig.~\ref{Fig:spinning_dust} along with the spinning dust emissivity of the anomalous emission observed with the VSA~\citep{Tibbs:10}. Region E was chosen as it corresponds to the location of the peak anomalous emission detected by the VSA.

From Fig.~\ref{Fig:spinning_dust} it is clear that the spinning dust models for each of the phases of the ISM are not consistent with the VSA measurement, with the model for the WNM representing the best agreement. Although we are using the values of $\chi_\mathrm{ISRF}$ that we have derived in this analysis, we believe that the reason for this lack of agreement is because the other idealised values for the environments are not physically representative of the region. For instance, the idealised value of $\chi_\mathrm{ISRF}$ for the~\citet{DaL:98} PDR model is 3$\times$10$^{3}$, which is clearly not consistent with the values of $\chi_\mathrm{ISRF}$ obtained in this work~(see Table~\ref{Table:final_params}), suggesting that the other idealised environmental values are also not appropriate. The result for the other regions of anomalous emission is similar, implying that to fit the observations we need more constraints on the other parameters within \textsc{spdust}, rather than using idealised values.

\begin{figure}
\begin{center}
\includegraphics[angle=0,scale=0.35, viewport=77 50 700 570]{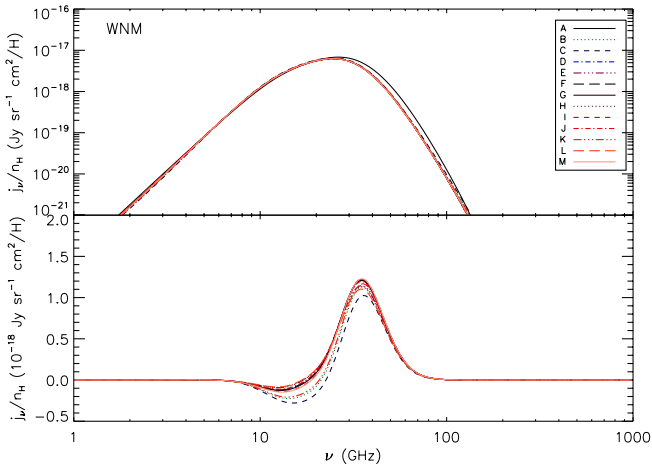} \\
\caption{Spinning dust curves computed with \textsc{spdust} for all thirteen regions using the $\chi_\mathrm{ISRF}$ values constrained with \textsc{dustem} and the idealised parameters for the WNM~(\textit{Top}). The difference in these spinning dust curves relative to region A~(\textit{Bottom}). This shows clearly, that even in regions with no anomalous emission, the modelled spinning dust emissivity is similar to regions in which we detect anomalous emission. This implies that these modelled curves cannot explain the observed emissivity variations, and once again highlights the complexity of this emission mechanism.}
\label{Fig:emiss_WNM}
\end{center}
\end{figure}

We also find that even in regions where we do not detect any anomalous emission, the modelled spinning dust emissivity is similar to regions where we do detect anomalous emission. This idea is shown in Fig.~\ref{Fig:emiss_WNM} where the modelled spinning dust emissivity for the WNM is plotted for all thirteen regions: remarkably, the computed spinning dust curves have very little dependence on $\chi_\mathrm{ISRF}$. This highlights that, although $\chi_\mathrm{ISRF}$ was found to play a significant role in the excitation process of the dust grains, the strength of the ISRF alone is not enough to accurately constrain the spinning dust emission and distinguish between regions with and without anomalous emission. This, in turn, once again emphasises the complex nature of this emission mechanism and demonstrates that other parameters, not yet constrained by this analysis, may play a bigger role.


\section{Conclusions}
\label{sec:conclusions}

This work represents the first attempt to characterise the anomalous emission by using IR observations  to constrain the dust properties. By reprocessing the original c2d \textit{Spitzer} photometric observations of the Perseus cloud we have produced maps of a substantially improved quality. Using these reprocessed maps and the 2MASS/NICER extinction map from the COMPLETE survey, in conjunction with the dust emission model \textsc{dustem}, we have been able to constrain the abundances of the PAHs and VSGs relative to the BGs, the strength of the ISRF relative to the~\citet{Mathis:83} solar neighbourhood value and the thermal equilibrium temperature of the BGs, resulting in parameter maps of the region at 7~arcmin angular scales. The Y$_\mathrm{PAH}$, Y$_\mathrm{VSG}$ and $\chi_\mathrm{ISRF}$ parameter maps represent the first of their kind for the Perseus cloud.

We found that there is a significant increase in $\chi_\mathrm{ISRF}$, and consequently T$_\mathrm{dust}$, in regions of anomalous emission. This result can be interpreted in terms of the spinning dust hypothesis, with the anomalous emission possibly being produced by electric dipole emission from spinning dust grains excited by an increase in photon~--~grain interactions. This result also suggests that the spinning dust excitation mechanism plays a vital role in producing the anomalous emission observed with the VSA and as such the emission observed with the VSA is likely originating from the PDR surrounding the H\textsc{ii} gas in G159.6--18.5, and the open cluster IC~348. 

The results of our analysis also suggest that, due to the difference in the observed angular scales, the anomalous emission detected with the VSA is different to that detected by the~\citet{Planck_Dickinson:11}, which would appear to explain why we find that $\chi_\mathrm{ISRF}$ plays an important role for the anomalous emission observed with the VSA, but not for the \textit{Planck} observations at degree angular scales.

In this work we demonstrate the complex nature of the anomalous emission. With more IR observations becoming available~(e.g. \textit{Herschel} data), it will be possible to constrain additional dust properties to even higher angular resolution. Combining this with future high resolution microwave observations will help to identify, in even more detail, the physical conditions in which anomalous emission originates. This type of analysis creates a new perspective in the field of anomalous emission studies, and represents a powerful new tool for constraining spinning dust models.


\section*{Acknowledgments}

We thank the anonymous referee for their careful reading of this paper and for providing useful comments. This work is based in part on archival data obtained with the \textit{Spitzer Space Telescope}, which is operated by the Jet Propulsion Laboratory, California Institute of Technology under a contract with NASA. Support for this work was provided by an award issued by JPL/Caltech. This work has been done within the framework of a NASA/ADP ROSES-2009 grant, \# 09-ADP09-0059. CD acknowledges support from an STFC Advanced Fellowship and an ERC IRG grant under the FP7. SC acknowledges support from FONDECYT grant 1100221, and from the Chilean Center for Astrophysics FONDAP 15010003.




\bsp 

\label{lastpage}

\end{document}